\documentclass[a4paper,11pt]{article}

\usepackage{bm}
\usepackage{anysize}
\usepackage{amsthm}
\usepackage{amssymb}
\usepackage{amsmath}
\usepackage{fancybox}
\usepackage{bbm}
\usepackage{amsfonts}
\usepackage[T1]{fontenc}
\usepackage{latexsym}
\usepackage{makeidx}
\usepackage{cite}
\usepackage[numbers]{natbib}

\usepackage[pdftex]{graphicx}
\usepackage[pdftex]{color}

\usepackage[all]{xy}

\usepackage{stackengine}
\stackMath

\newtheorem{teo}{Theorem}
\newtheorem{lem}{Lemma}

\newtheorem{cor}{Corollary}
\newtheorem{defi}{Definition} 

\newcommand*{\m}[1]{\underline{#1}}

\newcommand{\fd}{\rightarrow}

\newcommand{\inc}{\subset}

\newcommand{\iso}{\cong}

\newcommand{\al}{\alpha}
\newcommand{\be}{\beta}
\newcommand{\lan}{\lambda}
\newcommand{\fhi}{\varphi}
\newcommand{\del}{\delta}
\newcommand{\Del}{\Delta}
\newcommand{\gam}{\gamma}
\newcommand{\Gam}{\Gamma}

\newcommand{\Om}{\Omega}

\newcommand{\Z}{\mathbb{Z}}
\newcommand{\N}{\mathbb{N}}

\newcommand{\R}{\mathbb{R}}
\newcommand{\C}{\mathbb{C}}

\newcommand{\E}{\mathbb{E}}

\newcommand{\Sa}{\mathbb{S}}

\newcommand{\pa}{\partial}

\newcommand{\sgn}{\mbox{sgn}}

\newcommand{\p}{\grave{}}

\newcommand{\el}{\ell}

\newtheorem{remark}{Remark}[section]

\def\pf{\par\noindent {\em Proof.}~\par\noindent}

\def\max{\mathop{\mbox{\normalfont max}}\limits}
\def\lim{\mathop{\mbox{\normalfont lim}}\limits}

\def\pf{\par\noindent {\em Proof. }}

\def\pa{\partial}

\begin{document}

\date{}

\title{Generalized Cauchy-Kovalevskaya extension and plane wave decompositions in superspace}
\small{
\author
{Al\'i Guzm\'an Ad\'an}
\vskip 1truecm
\date{\small  Clifford Research Group, Department of Electronics and Information Systems,\\
 Faculty of Engineering and Architecture, Ghent University, Krijgslaan 281, 9000 Gent, Belgium. \\
{\tt Ali.GuzmanAdan@UGent.be}}

\maketitle

\begin{abstract} 
{The aim of this paper is to} obtain a generalized CK-extension theorem in superspace for the bi-axial Dirac operator $\pa_{\bf x} +\pa_{\bf y}$. In the classical commuting case, this result can be written as a power series of Bessel type of certain differential operators acting on a single initial function. In the superspace setting, novel structures appear in the cases of negative even superdimensions. In these cases, the CK-extension depends on two initial functions on which two power series of differential operators act. These series are not only of Bessel type but they give rise to an additional structure in terms of Appell polynomials. This pattern also is present in the structure of the Pizzetti formula, which describes integration over the supersphere in terms of differential operators. We make this relation explicit by studying the decomposition of the generalized CK-extension into plane waves integrated over the supersphere. Moreover, these results {are applied to obtain a decomposition} of the Cauchy kernel in superspace into monogenic plane waves, which shall be useful for inverting the super Radon transform.


\noindent

\vspace{0.3cm}

\small{ }


\noindent
\textbf{}
\end{abstract}


\section{Introduction}
The Cauchy-Kovalevskaya extension theorem is a very well-known result (see e.g. \cite{MR2856647, MR1579652}), whose basic idea is to characterize solutions of suitable systems of PDE's by their restrictions (and the restrictions of some of their derivatives) to a submanifold of codimension one. When the PDE involved is the Cauchy-Riemann equation, it follows from this theorem that a holomorphic function in an appropriate region of the complex plane is completely determined by its restriction to the real axis. This extension principle for holomorphic functions has been elegantly extended to higher dimensions in the framework of Clifford analysis, which constitutes a higher dimensional generalization of holomorphic function theory in the complex plane, and a refinement of harmonic analysis, see e.g.\ \cite{MR1169463, MR697564, MR1130821}.

{Clifford analysis focusses on the study of monogenic functions, i.e.\ null-solutions of the Dirac operator $\pa_{\m{x}}=\sum_{j=0}^m e_j\pa_{x_j}$ or the Cauchy-Riemann operator $\pa_{x_0}+\pa_{\m{x}}$ in an open region of $\R^m$ or $\R^{m+1}$ respectively.}
Here $(e_1, \ldots, e_m)$ is an orthonormal basis in $\R^m$ underlying the construction of the Clifford algebra $\mathcal{C}_{m,0}$, and $\m{x}=\sum_{j=1}^m x_je_j$ is a vector variable defined in $\R^m$. In general, every monogenic function $f(x_0,\m{x})$, i.e. $(\pa_{x_0}+\pa_{\m{x}})f(x_0,\m{x}) =0$, is determined by its restriction to the hyperplane $x_0=0$. Conversely, any given real analytic function $f(\m{x})$ defined in a region of $\R^m$, has a unique monogenic extension $f(x_0,\m{x})$ called Cauchy-Kovalevskaya extension (CK-extension for short). 

Clifford analysis  also offers a framework for generalized CK-extensions that consider restrictions to submanifolds not only of codimension one but of arbitrary codimensions. In \cite{MR1169463}, a generalized CK-extension theorem was obtained for monogenic functions in an SO$(m)$-invariant domain $\Om\inc\R^{m+p}$ by considering their restrictions to $\R^p$.
This leads to a Taylor series for monogenic functions $f(\m{x},\m{y})$ such that $\m{y}\in \R^p$ is considered as a parameter and that we have a power series in the variable $\m{x}\in\R^m$.
{Putting $\Om_p=\Om\cap \R^p$, the generalized CK-extension theorem reads as follows (see \cite{MR1169463, MR2970974, MR3077647} for more details).}
\begin{teo}\label{Absolut} 
Let $f_{0}(\m{y})$ be (a Clifford-valued) analytic function in $\Omega_p$. Then there exists a unique sequence $\{f_j(\m{y})\}_{j=1}^\infty$ of analytic functions such that the series $f(\m{x},\m{y})=\sum_{j=0}^\infty \m{x}^j f_j(\m{y})
$
is convergent in a $(m+p)$-dimensional neighborhood of $\Omega_p$ and its sum $f$ is a monogenic function, i.e.\ $(\pa_{\m{x}} + \pa_{\m{y}}) f(\m{x},\m{y})=0$. The function $f_{0}(\m{y})$ is determined by the relation $f_{0}(\m{y})=f(\m{0},\m{y})$.
Furthermore, the sum $f$ is formally given by the expression
\begin{equation}\label{CKTrad}
f(\m{x},\m{y})=\Gamma\left(\frac{m}{2}\right) \left(\frac{|\m{x}|}{2}\sqrt{\Delta_{\m{y}}}\right)^{-\frac{m}{2}} \left(\frac{|\m{x}|}{2} \sqrt{\Delta_{\m{y}}}\, J_{\frac{m}{2}-1}\left(|\m{x}|\sqrt{\Delta_{\m{y}}}\right) +\frac{\m{x}\partial_{\m{y}}}{2} J_{\frac{m}{2}}\left(|\m{x}|\sqrt{\Delta_{\m{y}}}\right)\right)f_{0}(\m{y}),
\end{equation}
where $J_{\nu}$ is the Bessel function of the first kind of order $\nu$ and $\sqrt{\Delta_{\m{y}}}$ is the the formal square root of the Laplacian whit respect to $\m{y}$ (of which only even powers occur in the resulting series).
\end{teo}

The main goal of this paper is to tackle some interesting problems arising when considering the above generalized CK-extension 
in the superspace setting. Traditionally superspaces have been studied using methods from algebraic and differential geometry, see e.g.\ \cite{Berezin:1987:ISA:38130, Kostant:1975qe, MR2840967, MR2069561, MR565567, MR778559, MR574696}. In this paper, we use a more recent approach based on an extension of harmonic and Clifford analysis to superspace \cite{Bie2007, de2007clifford, MR2344451, 1751-8121-42-24-245204, MR2521367}. This extension is done by introducing suitable orthogonal and symplectic Clifford algebra generators, which allows to define a super Dirac operator, a super Laplace operator, and other important operators in superspace, thus constructing a representation of the Lie superalgebra $\mathfrak{osp}(1|2)$. Together with the traditional bosonic (commuting) variables $x_1,\ldots, x_m$, this extension considers an even number of fermionic (anti-commuting) variables due to the symplectic structure.  

%
%

In \cite{MR3989554},{a particular extension} {of Theorem \ref{Absolut}} to superspace was studied for monogenic superfunctions of two orthogonal supervector variables $F({\bf x},{\bf y})=\sum_{j=0}^\infty {\bf x}^j F_j({\bf y})$. {This study revealed a novel structure for the CK-extension (\ref{CKTrad}) when the supervector ${\bf x}$  is purely fermionic, i.e.\ it has no commuting coordinates. }
In this case, the resulting generalized Taylor series is not of Bessel type, as it is in the purely bosonic case (\ref{CKTrad}), but it gives rise to an additional structure in terms of Appell polynomials. Given these new insights, our first goal is {to completely characterize the generalized CK-extension theorem in superspace} for all possible superdimensions (see Theorem \ref{Gen_CK_extension}). {As we shall see,} it turns out that the Appell polynomial structure appears not only when ${\bf x}$ is purely fermionic. In general, the only requirement is that the superdimension of ${\bf x}$ is even and negative, or equivalently, that one of the powers ${\bf x}^j$ is monogenic. In the most interesting case, this leads to a CK-extension that combines two power series (of Appell polynomial type and Bessel type) of differential operators acting on two different initial functions respectively. 


The CK-extension formula (\ref{CKTrad}) is very related to the so-called Pizzetti formula, which expresses the integral over the unit sphere $\Sa^{m-1}\inc \R^m$ as a certain power series (also of Bessel type) of the Euclidean Laplace operator acting on the integrand, see \cite{Pizz}. This relation is explicitly realized (via the Funk-Hecke theorem) by means of the plane wave decomposition 
\begin{equation}\label{PWTrad}
f(\m{x},\m{y}) = \left(\frac{1}{\sigma_m} \int_{\Sa^{m-1}} \exp\left(\langle \m{w}, \m{x}\rangle \, \m{w}  \, \pa_{\m{y}}\right) \, dS_{\m{w}} \right) f_0(\m{y}),
\end{equation}
where $f(\m{x},\m{y})$ is defined as in Theorem \ref{Absolut},  $\langle \m{w}, \m{x}\rangle$ denotes the Euclidean inner product in $\R^m$, and $\sigma_m$ is the surface area of the unit sphere $\Sa^{m-1}$. 
{Formula (\ref{PWTrad}) expresses the Bessel type differential operator from Theorem \ref{Absolut} as an integral over $\Sa^{m-1}$ of a much simpler integrand of plane wave type.}
The above-mentioned structure of the generalized CK-extension formula in superspace shows remarkable differences with the classical case, {which makes} 
 the connection with the Pizzetti formula in this setting (see \cite{MR2539324, MR2344451}) very interesting. Our second goal is to study this connection by extending the above plane wave decomposition (\ref{PWTrad}) to superspace.


Again, in the cases of negative and even superdimension, we find that the extension of formula (\ref{PWTrad}) fails to preserve its classical form. In these cases, one encounters problems with the definition of the normalized integral over the supersphere since the supersphere area $\sigma_{-2k}$ vanishes, see \cite{MR2539324, Guz_Somm5}. However, the notion of normalized integral can still be defined for a certain class of superfunctions by considering a limit case of the Pizzetti formula (see Section \ref{NISS}).  The resulting Pizzetti series reduces, in this case, to a finite power series (with the form of an Apple polynomial) of the super Laplace operator acting on the integrand.  These coincidences in the structure of Pizzetti's formulas and the generalized CK-extension make it possible to decompose the latter into plane waves for all possible superdimensions (see {Theorem \ref{PWDecompCK}}).

In the purely bosonic case, Theorem \ref{Absolut} has important applications in proving plane wave decompositions of monogenic functions and, in particular, of the Cauchy kernel which is given by a fundamental solution of the Cauchy-Riemann operator $\pa_{x_0}+\pa_{\m{x}}$. This decomposition 
reads as follows {for $x_0\neq 0$}
\begin{equation}\label{PWCKTrad}
\frac{1}{\sigma_{m+1}}  \frac{x_0-{\m{x}}}{|x_0-{\m{x}}|^{m+1}} = 
\begin{cases}
\sgn(x_0) \frac{(-1)^{\frac{m}{2}}(m-1)!}{2 (2\pi)^m} \int_{\Sa^{m-1}} (\langle{\m{x}},{\m{w}}\rangle - x_0 {\m{w}})^{-m}\, dS_{\m{w}}, &  \mbox{ for } m \mbox{ even}, \\[+.2cm]
 \frac{(-1)^{\frac{m+1}{2}}(m-1)!}{2 (2\pi)^m} \int_{\Sa^{m-1}} (\langle{\m{x}},{\m{w}}\rangle - x_0 {\m{w}})^{-m} {\m{w}}\, dS_{\m{w}}, &  \mbox{ for } m \mbox{ odd}.
\end{cases}
\end{equation}
Examining the boundary values in the above expression for $x_0\fd 0^{\pm}$ yields a plane wave decomposition of the delta distribution in $\R^m$, which allows to reconstruct the classical Radon transform from a Clifford analysis perspective, see e.g.\ \cite{MR3167578, MR985370, MR1169463}. See also e.g.\ \cite{MR0166596, MR573446} for a classical treatment. These plane wave decompositions have been also used to obtain a Bony-type twisted Radon decomposition of the delta distribution in \cite{MR3167578}, which has applications in the theory of microlocalization and analytical wave front sets for distributions, see also \cite{MR1026013}. 

The final purpose of this paper is to extend the plane wave decompositions (\ref{PWCKTrad}) to the superspace setting as an application of the generalized CK-extension theorem.
To that end, we prove first that certain CK-extensions $F(x_0,{\bf x})= \sum_{j=0}^\infty {\bf x}^j F_j(x_0)$ can be decomposed into monogenic plane waves of the form $g(\langle{\bf x},{\bf w}\rangle - x_0 {\bf w} )$, where $g$ is a holomorphic complex function (see Lemmas \ref{Int_PW} and \ref{NorInt_PW}). This is the case for the super Cauchy kernel if we choose $g(z)=z^{-M}$, where $M\in\Z$ is the superdimension of ${\bf x}$. As in the previous analysis, when $M$ is even and negative, this plane wave decomposition no longer resembles the structure of (\ref{PWCKTrad}) (see Theorem \ref{PWDCK}). {These differences with the classical case are expected to produce new insights in the corresponding plane wave decomposition of the delta distribution in superspace. Consequently, this should have an impact on the
problem of inverting}
the Radon transform in superspace introduced in the papers \cite{MR2344451, MR2539324, MR2422641}, and for which no inversion formula is known yet.} These two problems will be the topic of future work. 

Summarizing, in this paper we solve the following problems:
 \begin{itemize}
\item {\bf P1:} obtain a generalized CK-extension theorem in superspace for all possible superdimensions
\item {\bf P2:} find a general decomposition of the generalized CK-extension into plane waves
\item {\bf P3:} provide a plane wave decomposition of the super Cauchy kernel.
\end{itemize}

{The paper is organized as follows. In Section \ref{Prem}, we give a short introduction on harmonic and Clifford analysis in superspace focusing on the notions needed in the sequel. In section \ref{NISS}, we introduce the notion of normalized integral over the supersphere (in negative even superdimensions) as a limit case of the classical Pizzetti formula. Moreover, we study some of its basic properties needed for the subsequent plane wave decompositions. Section  \ref{CKESect} is fully devoted to the proof of the generalized CK-extension theorem in superspace, solving thus ${\bf P1}$. In section \ref{PWDSect}, we show how to decompose the generalized CK-extension in terms of (normalized) integrals of plane waves over the supersphere, providing a solution to ${\bf P2}$.  In Section \ref{IMPWSect}, we study decompositions of certain generalized CK-extensions into plane waves constructed out of holomorphic functions. Finally, in Section \ref{PWDCKSect}, we use the previous results to obtain a plane wave decomposition of the super Cauchy kernel, which solves ${\bf P3}$.



\section{Preliminaries}\label{Prem}

\noindent Consider $m$ commuting (bosonic) variables $x_1,\ldots, x_m$ and $2n$ anti-commuting (fermionic) variables $x\p_1, \ldots, x\p_{2n}$ in a purely symbolic way, i.e.\ $x_jx_k = x_kx_j$, $x\p_j x\p_k = -x\p_k x\p_j$ and $x_jx\p_k = x\p_k x_j$. They give rise to the supervector variable
\[{\bf x}=(\underline{x},\underline{x\p})=\left(x_1,\ldots, x_m,x\p_1,\ldots, x\p_{2n}\right).\]
The variables $x_1,\ldots, x_m$ are generators of the polynomial algebra $\R[x_1,  \ldots, x_m]$ while $x\p_1, \ldots, x\p_{2n}$ generate a Grassmann algebra $\mathfrak{G}_{2n}$. We denote by $\mathfrak{G}^{(ev)}_{2n}$ and $\mathfrak{G}^{(odd)}_{2n}$ the subalgebras of even and odd elements of $\mathfrak{G}_{2n}$ respectively. All the variables together generate the algebra of super-polynomials
\[ \mathcal{P}:= \mbox{Alg}_{\R}(x_1,  \ldots, x_m, x\p_1, \ldots, x\p_{2n})= \R[x_1,  \ldots, x_m] \otimes \mathfrak{G}_{2n}.\]
The bosonic and fermionic partial derivatives $\pa_{x_j}=\frac{\pa }{\pa x_j}$, $\pa_{x\p_j}=\frac{\pa }{\pa x\p_j}$ are defined as endomorphisms on $\mathcal P$ by the  relations 
\begin{equation}\label{ParDer1}
\begin{cases} \pa_{x_j}[1]=0,\\
 \pa_{x_j} x_k- x_k \pa_{x_j}=\del_{j,k},\\
 \pa_{x_j} x\p_k=x\p_k \pa_{x_j}, \;\; 
 \end{cases}
\hspace{.5cm} 
\begin{cases} \pa_{x\p_j}[1]=0,\\
\pa_{x\p_j} x\p_k+ x\p_k \pa_{x\p_j}=\del_{j,k},\\
\pa_{x\p_j} x_k=x_k\pa_{x\p_j}, 
\end{cases}
\end{equation} 
which can be recursively applied  for both left and right actions.

Associated to these variables we consider the flat supermanifold $\R^{m|2n}=\left(\R^m, \mathcal{O}_{\R^{m|2n}}\right)$ where $\mathcal{O}_{\R^{m|2n}}$ is the structure sheaf that maps every open subset $\Om\inc \R^m$ into the graded algebra $C^\infty(\Om) \otimes {\mathfrak G}_{2n}$ of smooth functions in $\Om$ with values in the Grassmann algebra ${\mathfrak G}_{2n}$. The partial derivatives $\pa_{x_j}$, $\pa_{x\p_j}$ extend from $\mathcal P$ to $C^\infty(\R^m)\otimes \mathfrak{G}_{2n}$ by density. 


\noindent Let us rewrite the supervector variable ${\bf x}$  as
\[{\bf x}=\underline{x}+\underline{x\p}=\sum_{j=1}^mx_je_j+\sum_{j=1}^{2n}x\p_j e\p_j,\]
where $e_1,\ldots,e_m,e\p_1,\ldots,e\p_{2n}$ is the standard homogeneous  basis of the graded vector space $\R^{m,2n}=\R^{m,0}\oplus \R^{0,2n}$. Here we have denoted by $\underline{x}= \sum_{j=1}^m x_je_j$ and $\underline{x\p}=\sum_{j=1}^{2n}x\p_je\p_j$ the so-called bosonic and fermionic projections of ${\bf x}$ respectively. We consider an orthosymplectic metric in $\R^{m,2n}$, giving rise to the super Clifford algebra $\mathcal C_{m,2n}:=\mbox{Alg}_\R(e_1,\ldots,e_m,e\p_1,\ldots,e\p_{2n})$ governed by the multiplication rules 
\begin{align}\label{CommRules}
e_je_k+e_ke_j=-2\del_{j,k}, \hspace{.3cm} e_je\p_k+e\p_ke_j=0, \hspace{.3cm} e\p_je\p_k-e\p_ke\p_j=g_{j,k},
\end{align}
where $g_{j,k}$ is a symplectic form defined by
\[g_{2j,2k}=g_{2j-1,2k-1}=0, \hspace{.5cm} g_{2j-1,2k}=-g_{2k,2j-1}=\del_{j,k}, \hspace{.5cm} j,k=1,\ldots,n.\]


\noindent In this case the {\it inner product} of two supervectors ${\bf x}$ and ${\bf y}$ is given by
\[\langle{\bf x}, {\bf y}\rangle :=-\frac{1}{2}( {\bf x}{\bf y} + {\bf y}{\bf x})= \langle{\m x}, {\m y}\rangle + \langle{\m x\p}, {\m y\p}\rangle=\sum_{j=1}^mx_jy_j  -\frac{1}{2}\sum_{j=1}^n(x\p_{2j-1}y\p_{2j}-x\p_{2j}y\p_{2j-1}).\]


\noindent The {\it generalized norm squared} of the supervector ${\bf x}$ is thus defined by
\begin{equation}\label{NormSq}
|{\bf x}|^2=  \langle{\bf x}, {\bf x}\rangle = -{\bf x}^2=\sum_{j=1}^m x_j^2 - \sum_{j=1}^n x\p_{2j-1} x\p_{2j}.
\end{equation}
Observe that the fermionic vector variable $\m{x}\p$ is nilpotent. Indeed, its norm squared satisfies 
\[\underline{x}\p^{2n}=n! \,x\p_1x\p_2\cdots x\p_{2n-1}x\p_{2n},\]
which is the element of maximal degree in $\mathfrak{G}_{2n}$.


Functions in $C^\infty(\Om)\otimes \mathfrak{G}_{2n}$ {(often called superfunctions)} can be explicitly written as 
\begin{equation}\label{SupFunc}
F({\bf x})=F(\underline{x},\underline{x\p})=\sum_{A\inc \{1, \ldots, 2n\}} \, F_A(\underline{x})\, \underline{x}\p_A, 
\end{equation}
where $F_A(\underline{x})\in C^\infty(\Om)$ and  $ \underline{x}\p_A=x\p_{j_1}\ldots x\p_{j_k}$ with $A=\{j_1,\ldots, j_k\}$, $1\leq j_1< \ldots< j_k\leq 2n$.  Every superfunction can be written as the sum $F({\bf x})= F_0(\underline{x})+ {\bf F}(\underline x, \underline{x}\p)$ where the  real-valued function $F_0(\underline{x})=F_\emptyset(\underline{x})$ is called the {\it body} $F$, and ${\bf F}=\sum_{|A|\geq 1} \, F_A(\underline{x})\,\underline{x}\p_A$ is the {\it nilpotent part} of $F$. Indeed, it is clearly seen that  ${\bf F}^{2n+1}=0$. We shall also consider the space of analytic superfunctions $\mathcal{A}(\Om)\otimes \mathfrak{G}_{2n}$, i.e.\ functions of the form (\ref{SupFunc}) where each of the $F_A$'s belongs to the space $\mathcal{A}(\Om)$ of real analytic functions on $\Om\inc \R^m$. When considering the aforementioned Clifford generators, these spaces of superfunctions extend to the spaces of Clifford-valued superfunctions $C^\infty(\Om)\otimes \mathfrak{G}_{2n}\otimes \mathcal{C}_{m,2n}$ and $\mathcal{A}(\Om)\otimes \mathfrak{G}_{2n}\otimes \mathcal{C}_{m,2n}$ respectively. 


\noindent The  bosonic and fermionic Dirac operators are defined by
\[\pa_{\underline x}=\sum_{j=1}^m e_j\pa_{x_j}, \hspace{1cm} \pa_{\underline x\p}=2\sum_{j=1}^n \left(e\p_{2j}\pa_{x\p_{2j-1}}-e\p_{2j-1}\pa_{x\p_{2j}}\right),\]
giving rise to the left and right super Dirac operators (super-gradient)
$\pa_{\bf x} \cdot =\pa_{\underline x\p}\cdot -\pa_{\underline x}\cdot$ and $\cdot\pa_{\bf x}  =- \cdot\pa_{\underline x\p} - \cdot \pa_{\underline x}$ respectively. As in the classical setting, the action of $\pa_{\bf x}$ on the vector variable ${\bf x}$ results in the superdimension
\[M:=\pa_{\bf x}[{\bf x}]=[{\bf x}]\pa_{{\bf x}}=\pa_{\underline{x\p}}[\underline{x\p}]-\pa_{\underline{x}}[\underline{x}]=m-2n.\]
The action of the super Dirac operator on any positive power of ${\bf x}$ is given by 
\begin{align}\label{ActDirOnPow}
\pa_{\bf x}[{\bf x}^j]&=c(M,j){\bf x}^{j-1}, & & \mbox{where}& c(M,j)&=\begin{cases}  j, & \mbox{ if } j \mbox{ is even}, \\ M+j-1, & \mbox{ if } j \mbox{ is odd}. \end{cases}
\end{align}

Given an open set $\Om\inc \R^m$, a superfunction $F\in C^\infty(\Om)\otimes \mathfrak{G}_{2n}\otimes \mathcal C_{m,2n}$ is said to be (left) {\it monogenic} if $\pa_{\bf x}[F]=0$. As the super Dirac operator factorizes the super Laplace operator:
\[\Del_{m|2n}=-\pa_{\bf x}^2=\sum_{j=1}^{m} \pa^2_{x_j}-4\sum_{j=1}^n \pa_{x\p_{2j-1}}\pa_{x\p_{2j}},\]
monogenicity also constitutes  a refinement of harmonicity in superanalysis. {Whenever necessary, we shall also use the alternative notation $\Del_{\bf x}$ for the Laplacian $\Del_{m|2n}$ to explicitly indicate that we are differentiating with respect to the vector variable ${\bf x}$.}

The super Euler operator is defined by
\[
\E=\sum_{j=1}^m x_j\pa_{x_j} + \sum_{j=1}^{2n} x\p_j\pa_{x\p_j}.
\]
We denote by $\N_0:=\{0\}\cup\N$ the set of non-negative integers. Homogeneous super-polynomials of degree $j\in \N_0$ are eigenfunctions of the super Euler operator with eigenvalue $j$. We denote the space of homogeneous super-polynomials of degree $j\in \N_0$ as $\mathcal{P}_j= \{R\in \mathcal{P} : \E[R]=j\,R\}$, which allows for the decomposition
\[
\mathcal{P} = \bigoplus_{j=0}^\infty \mathcal{P}_j.
\]
An element $R\in \mathcal P$ is called a {spherical harmonic of degree $j$} if it satisfies 
\[
\Del_{m|2n}[R]=0,  \;\;\;\; \mbox{ and }  \;\;\;\; \E[R]=jR,  \;\; \mbox{(i.e. $R\in \mathcal{P}_j$)}.
\]
The space of all spherical harmonics of degree $j$ is denoted by $\mathcal{H}_j$. 

The operators $\Del_{m|2n}$, ${\bf x}^2$ and $\E$ satisfy the canonical commutation relations of $\mathfrak{sl}_2$   (see e.g.\ \cite{MR2344451})
\begin{align}\label{sl2}
\left[\frac{\Del_{m|2n}}{2}, \frac{-{\bf x}^2}{2}\right] &= \E + \frac{M}{2}, & \left[\frac{\Del_{m|2n}}{2}, \E + \frac{M}{2}\right] &= {\Del_{m|2n}}, & \left[\frac{-{\bf x}^2}{2}, \E + \frac{M}{2}\right] &= {{\bf x}^2}.
\end{align}
Similarly to classical Clifford analysis, $\pa_{\bf x}$ and ${\bf x}$ satisfy the commutation rules of $\mathfrak{osp}(1|2)$.  Thus the same computation rules of classical Clifford and harmonic analysis can be transferred to the superspace setting by substituting the Euclidean dimension $m$ by the superdimension $M$. In particular, the following Lemma can be proved using formulae (\ref{sl2}) iteratively, see \cite{MR2344451}.

\begin{lem}\label{LemLap}
Let $R_{2j} \in \mathcal{P}_{2j}$. Then the following identity holds for any superdimension $M\in\Z$,
\[
\Del_{m|2n}^{j+\el}[{\bf x}^{2\el} R_{2j}] = (-1)^{\el} 4^{\el} \frac{(j+\el)!}{j!} \frac{\Gam\left(j+\el+\frac{M}{2}\right)}{\Gam\left(j+\frac{M}{2}\right)}\, \Del_{m|2n}^{j}[R_{2j}],
\]
with $\frac{\Gam\left(\el+q\right)}{\Gam\left(q\right)} := (q)_\el = q(q+1) \cdots (q+\el-1)$ for all $q\in \R$, $\el \in \N$.
\end{lem}
More details on the theory of monogenic and harmonic superfunctions can be found for instance in \cite{MR2344451, 1751-8121-42-24-245204, MR2386499, MR2405887, MR3375856}.

Given a set of superfunctions $\{F_j\}_{j\in\N_0} \inc C^\infty(\Om)\otimes \mathfrak{G}_{2n}$ with $F_j({\bf x}) = \sum_{A\inc \{1, \ldots, 2n\}} \, F_{A,j}(\underline{x})\, \underline{x}\p_A$,
we define 
\begin{equation}\label{series}
\sum_{j=0}^\infty F_j ({\bf x}) =  \sum_{A\inc \{1, \ldots, 2n\}} \left( \sum_{j=0}^\infty  F_{A,j}(\underline{x}) \right) \underline{x}\p_A.
\end{equation}
Thus we say that the series (\ref{series}) converges uniformly, absolutely or normally in $E\inc \Om$ if each of the series $\sum_{j=0}^\infty  F_{A,j}(\underline{x})$ converges uniformly, absolutely or normally respectively in $E\inc \Om$ for all $A\inc \{1, \ldots, 2n\}$. 

We now proceed to define series of Clifford-valued superfunctions. First observe that the algebra $\mathcal{C}_{m,2n}$ is infinite dimensional if $n\neq 0$ with basis
\[
B = \{e_1^{\gam_1} \ldots e_m^{\gam_m} e\p_2^{\,\al_1} \ldots e\p_{2n}^{\,\al_n} e\p_1^{\,\be_1} \ldots e\p_{2n-1}^{\,\be_n} : \gam_j \in \{0,1\}, \al_j, \be_j\in \N_0 \}. 
\]
Consider now 
a set of superfunctions $\{F_j\}_{j\in\N_0} \inc C^\infty(\Om)\otimes \mathfrak{G}_{2n} \otimes \mathcal{C}_{m,2n}$ such that they all are spanned by a finite set $\{a_1, \ldots, a_s\}\inc B$ of basis elements, i.e.\ for each $j\in \N_0$ we have
\[
 F_j ({\bf x})=  \sum_{r=1}^s F_{r}^{(j)}( { \bf x})\,  a_r, \;\;\;\;  F_{r}^{(j)}( { \bf x}) \in C^\infty(\Om)\otimes \mathfrak{G}_{2n}.
\] 
For such a set of superfunctions we define 
\begin{equation}\label{SerClifSF}
\sum_{j=0}^\infty F_j ({\bf x})=  \sum_{r=1}^s \left(\sum_{j=0}^\infty F_{r}^{(j)}( { \bf x})\right)  a_r,
 \end{equation}
where the innermost series $\sum_{j=0}^\infty F_{r}^{(j)}( { \bf x})$, $r=1, \ldots, s$, are understood in the sense of (\ref{series}).

%

A form of producing interesting even superfunctions is by considering finite Taylor expansions of real-valued functions, see e.g.\ \cite{Berezin:1987:ISA:38130}. 
\begin{defi}\label{CompSupFunct}
Consider a function $F \in C^\infty(E)$ where $E$ is an open region of $\R^\ell$, and $\el$ even superfunctions $a_j({\bf x})\in C^\infty(\Om)\otimes {\mathfrak G}^{(ev)}_{2n}$, $j=1,\ldots, \ell$. We expand every $a_j$ as the sum of its body and its nilpotent part, i.e.\ $a_j({\bf x})=[a_j]_0(\underline{x})+{\bf a_j}({\bf x})$. If the domain $E\subseteq \R^\ell$ contains the image of the function $([a_1]_0, \ldots, [a_\el]_0)$, we define the composed superfunction $F\left(a_1({\bf x}), \ldots,a_\ell({\bf x})\right) \in C^\infty(\Om)\otimes {\mathfrak G}_{2n}$ by means of the Taylor expansion as
\begin{equation}\label{Tay_Ser}
F(a_1, \ldots, a_\ell)= \sum_{k_1,\ldots, k_\ell\geq 0} \frac{F^{(k_1,\ldots, k_\ell)}\left([a_1]_0, \ldots, [a_\ell]_0\right)}{k_1! \cdots k_\ell!}  {\bf a_1}^{k_1} \cdots {\bf a_\ell}^{k_\ell}.
\end{equation}
\end{defi}

\begin{remark}
Note that the series in the above definition of $F(a_1, \ldots, a_\ell)$ is finite in view of the nilpotency of ${\bf a_j}({\bf x})$. Moreover, it is clear that Definition \ref{CompSupFunct} can be used for functions that are not $C^\infty$ as long as all the derivatives appearing in the formula exist.
\end{remark}

If the function $F$ is analytic, straightforward calculations show that the above expression is independent of the splitting of the even superfunctions $a_j$'s.

\begin{lem}\label{Lem1}
{Let  $F \in \mathcal{A}(E)$ be an analytic function in the open region $E\inc \R^\ell$, and let  $a_1, \ldots, a_\el, b_1, \ldots, b_\el\in C^\infty(\Om)\otimes {\mathfrak G}^{(ev)}_{2n}$. Suppose that the image of the function $([a_1]_0, \ldots, [a_\el]_0)$ is contained in $E$ and that $([a_1]_0+[b_1]_0, \ldots, [a_\el]_0+[b_\el]_0)$ is always in the region of convergence of the Taylor series of $F$ around  $([a_1]_0, \ldots, [a_\el]_0)$. Then one has }
\[F(a_1+b_1, \ldots, a_\ell+b_\el)= \sum_{k_1,\ldots, k_\ell\geq 0} \frac{F^{(k_1,\ldots, k_\ell)}\left(a_1, \ldots, a_\ell\right)}{k_1! \cdots k_\ell!}  {b_1}^{k_1} \cdots {b_\ell}^{k_\ell},\]
where each of the superfunctions $F^{(k_1,\ldots, k_\ell)}\left(a_1, \ldots, a_\ell\right)$ is understood in the sense of (\ref{Tay_Ser}).
\end{lem}
The expansion (\ref{Tay_Ser}) is used to define arbitrary real powers of even superfunctions. Let $p\in \R$ and $a=a_0+{\bf a}\in C^\infty(\Om)\otimes \mathfrak{G}_{2n}^{(ev)}$, then for $a_0>0$ we define  
\begin{equation}\label{GenPow}
a^p=\sum_{j=0}^{n}\;\frac{{\bf a}^j}{j!}\; \frac{\Gam(p+1)}{\Gam(p-j+1)} \,a_0^{p-j}. 
\end{equation}
If $m\neq 0$, one can follow this idea to define the {norm} of the supervector variable ${\bf x}$. Indeed, its norm squared $-{\bf x}^2$ is an even smooth superfunction with non-negative body $|\m{x}|^2=\sum_{j=1}^m x_j^2$, see (\ref{NormSq}). {Hence, the norm of ${\bf x}$ is defined as}
\[|{\bf x}|=(-{\bf x}^2)^{1/2}=\left(|\underline{x}|^2- \underline{x\p}^{\, 2} \right)^{1/2}=\sum_{j=0}^n \frac{(-1)^j \underline{x\p}^{\,2j}}{j!} \, \frac{\Gam(\frac{3}{2})}{\Gam(\frac{3}{2}-j)} |\underline{x}|^{1-2j}.\]

\section{{Normalized integral over the supersphere}}\label{NISS}
In this section we introduce the notion of normalized integral over the supersphere in the cases of negative even superdimensions following the ideas of \cite{MR2344451}. 
This is a key element to consider in the plane wave decomposition of the generalized CK-extension, see Section \ref{PWDSect}. 

The analogue in superspace of the classical integral $\int_{\R^m}\; dV_{\underline x}$ in $\R^m$ is given by
\[\int_{\R^{m|2n}} = \int_{\R^m} dV_{\underline{x}} \int_B=\int_B \int_{\R^m} dV_{\underline{x}},\]
where $dV_{\underline{x}}=d{x_1}\cdots d{x_m}$ the classical volume element in $\R^m$ and the integral over fermionic variables is given by the Berezin integral (see \cite{Berezin:1987:ISA:38130}), defined by 
\[\int_B=\pi^{-n} \, \pa_{x\p_{2n}}\cdots \pa_{x\p_{1}}=\frac{(-1)^n \pi^{-n}}{4^n n!} \pa_{\underline{x}\p}^{2n}.\]
In \cite{Guz_Somm5}, integration over general supermanifolds of codimension $0$ and $1$ was introduced by means of the action of the Heaviside and Dirac distributions respectively. In the case $m\neq 0$, the supersphere $\Sa^{m-1,2n}$ is algebraically defined by the relation ${\bf x}^2+1=0$. The classical integral over the unit sphere in $\R^m$ is extended to $\Sa^{m-1,2n}$ as (see also \cite{MR2344451, MR2539324})
\begin{equation}\label{IntSupSph}
 \int_{\Sa^{m-1,2n}} F({\bf x}) \, dS_{\bf x}= 2\int_{\R^m} \int_B \del({\bf x}^2+1) \,F({\bf x})\; dV_{\underline x},
 \end{equation}
where $\del({\bf x}^2+1)=\sum_{j=0}^n \frac{\underline{x}\p^{\,2j}}{j!} \;\del^{(j)}(1-|\underline{x}|^2)$ is the concentrated delta distribution on the supersphere. This extension is (up to a multiplicative constant) the unique $\mathfrak{osp}(m|2n)$-invariant integration operator over the supersphere that satisfies (see \cite{MR3060765})
\[\int_{\Sa^{m-1,2n}} |{\bf x}|^2 \, F({\bf x}) \, dS_{\bf x} = \int_{\Sa^{m-1,2n}} F({\bf x}) \, dS_{\bf x}.\]
This last property can be further generalized as follows.
\begin{lem}\label{Int_Lem1}
Let $f:\R\fd \R$ be a {smooth} function in a neighborhood of the point $x=1$.
Then for any $g\in C^\infty(E)\otimes \mathfrak{G}_{2n}$ with $\Sa^{m-1}\inc E \inc \R^m$ we have
\[\int_{\Sa^{m-1,2n}} f(|{\bf x}|) g({\bf x})\, dS_{\bf x} = f(1) \int_{\Sa^{m-1,2n}}  g({\bf x})\, dS_{\bf x}.\]
\end{lem}
\pf 
We recall that 
\[
\int_{\Sa^{m-1,2n}} f(|{\bf x}|) g({\bf x})\, dS_{\bf x} = 2\int_{\R^m} \int_B \del({\bf x}^2+1)\,f(|{\bf x}|) \, g({\bf x})\; dV_{\underline x}.
\]
Let us consider now the generalized function $G(t)=\del(1-t) f(t^{1/2})$, $t>0$. Then
\[
 \del({\bf x}^2+1)\,f(|{\bf x}|) = G(-{\bf x}^2) = G (|\m{x}|^2-\m{x}\p^{\,2}) = \sum_{j=0}^n \frac{\left(-\m{x}\p^{\,2}\right)^j}{j!} \, G^{(j)}(|\m{x}|^2).
\]
On the other hand, we have that $G(t)=\del(t-1) f(1)$ in the distributional sense, which implies that $G^{(j)}(t)=f(1)\del^{(j)}(t-1)$. 
Finally, using the identity $\del(-t)=\del(t)$, we obtain
\[
 \del({\bf x}^2+1)\,f(|{\bf x}|) = f(1)  \sum_{j=0}^n \frac{\left(-\m{x}\p^{\,2}\right)^j}{j!} \, \del^{(j)}(|\m{x}|^2-1) = f(1) \, \del(|\m{x}|^2-\m{x}\p^{\,2}-1) = f(1) \, \del({\bf x}^2+1),
\]
which proves the Lemma.
$\hfill\square$

In \cite{MR2539324}, it was proven that the integral (\ref{IntSupSph}) over $\Sa^{m-1,2n}$ reduces to the following Pizzetti formula when integrating over super-polynomials
\begin{equation}\label{PizzSS}
\int_{\Sa^{m-1,2n}} R({\bf x})\, dS_{\bf x}=\sum_{j=0}^\infty  \frac{2\pi^{M/2}}{2^{2j}\, j!\, \Gamma(j+M/2)} \Del_{m|2n}^j [R]\Big|_{{\bf x} =0} =\Phi_M(\Del_{m|2n})[R] \Big|_{{\bf x}=0},
\end{equation}
where $\displaystyle \Phi_M(z)=(2\pi)^{\frac{M}{2}} \frac{J_{\frac{M}{2}-1}(iz^{1/2})}{(iz^{1/2})^{\frac{M}{2}-1}}$ and $J_\nu$ is the Bessel function of first kind. In particular, one obtains that the surface area $ \sigma_{M}$ of  $\Sa^{m-1,2n}$ is given by $\sigma_{M}= \frac{2\pi^{\frac{M}{2}}}{\Gam\left(\frac{M}{2}\right)}$.

For $M=-2k$ and $m\neq 0$, the first $k+1$ terms in the above sum (\ref{PizzSS}) vanish and Pizzetti's formula reduces to 
\begin{equation}\label{Pizz-2k}
\int_{\Sa^{m-1,2n}} R({\bf x})\, dS_{\bf x}=\sum_{j=k+1}^\infty \frac{2\pi^{M/2}}{2^{2j}\, j!\, \Gamma(j+M/2)} \Del_{m|2n}^j [R]\Big|_{{\bf x}=0}.
\end{equation}
This implies that the integral of any polynomial of degree $\leq 2k+1$ vanishes. In particular, \[\sigma_{-2k} = \int_{\Sa^{m-1,2n}} 1 \, dS_{\bf x} =  \frac{2\pi^{-k}}{\Gam\left(-k\right)}=0.\]

In the purely fermionic case $m=0$, the supersphere cannot be defined as before since the {norm squared} ${\m{x}\p}^{\,2}$  of the vector variable $\m{x}\p$ is a nilpotent element in $\mathfrak{G}_{2n}$. Nevertheless, Pizzetti's formula (\ref{Pizz-2k}) provides a functional over $\mathfrak{G}_{2n}$ that can be regarded as {\it the integral over the supersphere} in this case. However, this extension is not very interesting since $\Del_{0|2n}^{n+1}=(-1)^{n+1} \pa_{\m{x\p}}^{2n+2}=0$,  which gives rise to the trivial functional $\int_{\Sa^{-1,2n}} \, dS_{\bf x} \equiv 0$. A more interesting functional is provided by the normalized integral. 


The normalized integral $\frac{1}{\sigma_{M}} \int_{\Sa^{m-1,2n}} F({\bf x})\, dS_{\bf x}$ is obviously well-defined if $M\notin -2\N_0$. On the other hand, if $M=-2k$ ($k\in \N_0$) and $\int_{\Sa^{m-1,2n}} F({\bf x})\, dS_{\bf x} \neq 0$, then the normalized integral is clearly undefined. However, for certain functions $F$ with a vanishing integral over the supersphere, it is possible to define a (non-vanishing) normalized integral $\frac{1}{\sigma_{-2k}} \int_{\Sa^{m-1,2n}} F({\bf x})\, dS_{\bf x}$. The idea behind this definition is as follows.

For a general superdimension $M$, the Pizzetti formula for the normalized integral of a polynomial $R$ of degree $\leq 2k+1$ reads as
\begin{align*}
\frac{1}{\sigma_{M}} \int_{\Sa^{m-1,2n}} R({\bf x})\, dS_{\bf x} &= \sum_{j=0}^k  \frac{\Gam(M/2)}{2^{2j}\, j!\, \Gamma(j+M/2)} \Del_{m|2n}^j [R]\Big|_{{\bf x}=0}.
\end{align*}
The coefficients  $\frac{\Gam(M/2)}{2^{2j}\, j!\, \Gamma(j+M/2)}$ ($j\leq k$) have a removable singularity at $M=-2k$. Indeed, when we see $M$ as a real (or complex) parameter, we have that $\lim_{M\fd -2k} \frac{\Gam(M/2)}{2^{2j}\, j!\, \Gamma(j+M/2)} = \frac{(-1)^j (k-j)!}{2^{2j}\, j!\,k!}$. This motivates the definition of the following functional on the space of polynomials $\oplus_{j=0}^{2k+1} \mathcal{P}_j$ 
in $\R^{m|2n}$ with superdimension $M=-2k$ (see also \cite{MR2344451})
\begin{align}\label{NormInt-2k}
\frac{1}{\sigma_{-2k}} \int_{\Sa^{m-1,2n}} R({\bf x})\, dS_{\bf x} := \frac{1}{k!}  \sum_{j=0}^k \frac{(k-j)!}{j!} \left(\frac{-\Del_{m|2n}}{4}\right)^j [R]\Big|_{{\bf x}=0} = \frac{1}{k!} \, P_k\left(\frac{-\Del_{m|2n}}{4}\right) [R]\Big|_{{\bf x}=0},
\end{align}
where $\{P_k\}_{k\in\N_0}$ is an Appell sequence of polynomials given by $P_k(z)=\sum_{j=0}^k \frac{(k-j)!}{j!} z^j$.
As mentioned before, this definition is particularly interesting in the purely fermionic case $m=0$ where the functional analogous to the integral  over the supersphere is trivial.

If $M=-2k$ and $m\neq 0$, we can use Lemma \ref{Int_Lem1} to extend the definition (\ref{NormInt-2k}) to functions of the form $f(|{\bf x}|) R({\bf x})$, with $R\in \oplus_{j=0}^{2k+1} \mathcal{P}_j$,
by setting
\begin{align}\label{NormIntExt}
\frac{1}{\sigma_{-2k}} \int_{\Sa^{m-1,2n}} f(|{\bf x}|) R({\bf x})\, dS_{\bf x} := f(1) \; \frac{1}{\sigma_{-2k}} \int_{\Sa^{m-1,2n}} R({\bf x})\, dS_{\bf x}.
\end{align}

\begin{remark}
The above extension is compatible with the original definition (\ref{NormInt-2k}). Indeed, if $R_{2j} \in \mathcal{P}_{2j}$ and $\el+j\leq k$, then by Lemma \ref{LemLap} one has
\begin{align*}
\frac{1}{\sigma_{-2k}} \int_{\Sa^{m-1,2n}} |{\bf x}|^{2\el} R_{2j}({\bf x})\, dS_{\bf x} &= \frac{(-1)^{j+\el}}{k!} \frac{(k-j-\el)!}{4^{j+\el} (j+\el)!}\, \Del_{m|2n}^{j+\el} \left[|{\bf x}|^{2\el} R_{2j}\right] \\
&= \frac{(-1)^j}{k!} \frac{(k-j)!}{4^j j!}\, \Del_{m|2n}^{j} \left[R_{2j}\right] \\
&= \frac{1}{\sigma_{-2k}} \int_{\Sa^{m-1,2n}} R_{2j}({\bf x})\, dS_{\bf x}.
\end{align*}
\end{remark}
For $M=-2n$ (i.e.\ $m=0$) the relation (\ref{NormIntExt}) holds if the sum of the degrees of $f$ and $R$ is no bigger than $2n$, {otherwise one may find cases where $fR \equiv 0$.} 
\begin{lem}\label{Int_Lem2}
Let $m=0$  and $\frac{1}{\sigma_{-2n}} \int_{\Sa^{-1,2n}}$ be defined as in (\ref{NormInt-2k}). Then, for $R_{2j}(\m{x}\p)\in \mathcal{P}_{2j}$ with $j\leq n$, one has
\begin{align*}
\frac{1}{\sigma_{-2n}} \int_{\Sa^{-1,2n}} \m{x}\p^{\,2\el} R_{2j}(\m{x}\p)\, dS_{\m{x}\p} = \begin{cases} \displaystyle (-1)^\el \; \frac{1}{\sigma_{-2n}} \int_{\Sa^{-1,2n}}  R_{2j}(\m{x}\p)\, dS_{\m{x}\p} & \el+j\leq n, \\[+.3cm]
0 & \el+j>n. \end{cases}
\end{align*}
\end{lem}

We shall make use of the Funk-Hecke theorem in superspace. In particular we will need the following particular case, see e.g. \cite{MR2683546, MR2344451} .
\begin{teo}\label{F-H_The}{\bf [Funk-Hecke]}
Consider $m\neq 0$ and let ${\bf x}, {\bf w}$ be independent super vector variables. Let $H_\el\in \mathcal{H}_\el$ and $j\in\N_0$, then
\[\int_{\Sa^{m-1,2n}} \langle {\bf x}, {\bf w} \rangle^j H_\el({\bf w})\, dS_{\bf w} = \al_{M,\el}[t^j] \, |{\bf x}|^{j-\el}\, H_\el({\bf x}),\]
where $\displaystyle \al_{M,\el}[t^j] = \begin{cases} \frac{j!}{(j-\el)!} \frac{2\pi^{\frac{M-1}{2}}}{2^\el} \frac{\Gam\left(\frac{j-\el+1}{2}\right)}{\Gam\left(\frac{M+j+\el}{2}\right)} & \mbox{ if } j+\el \mbox{ even and  } j\geq \el, \\ 0, & \mbox{ otherwise.}\end{cases}$
\end{teo}
A similar result was obtained in \cite{MR2344451} for the normalized integral (\ref{NormInt-2k}) in the case $m\neq 0$. The case $m=0$ can also be proved following almost an identical procedure to the one in \cite{MR2344451}. 
\begin{cor}\label{F-H_TheNor}{\bf [Funk-Hecke for the normalized integral]}
Let $M=-2k$ (including the case $m=0$) and $j+\el\leq 2k+1$. Then 
\[\frac{1}{\sigma_{-2k}} \int_{\Sa^{m-1,2n}} \langle {\bf x}, {\bf w} \rangle^j H_\el({\bf w})\, dS_{\bf w} = \al_{M,\el}^*[t^j] \, {\bf x}^{j-\el}\, H_\el({\bf x}),\]
where $\displaystyle \al^*_{M,\el}[t^j] = \begin{cases}   \frac{(-1)^j \pi^{-1/2}}{2^\el} \frac{\left(k-\frac{j+\el}{2}\right)!}{k!} \frac{j!}{(j-\el)!} \Gam\left(\frac{j-\el+1}{2}\right) & \mbox{ if } j+\el \mbox{ even and  } j\geq \el, \\ 0, & \mbox{ otherwise.}\end{cases}$
\end{cor}

\section{Generalized CK-extension theorem}\label{CKESect}
Now let us consider functions of the two independent supervector variables
\begin{align*}
{\bf x}&=\underline{x}+\underline{x\p}=\sum_{j=1}^m x_je_j+\sum_{j=1}^{2n}x\p_j e\p_j & &\mbox{and} & {\bf y}&=\underline{y}+\underline{y\p}=\sum_{j=1}^p y_je_{m+j}+\sum_{j=1}^{2q}y\p_j e\p_{2n+j},
\end{align*} 
where the elements $e_1, \ldots, e_m, e_{m+1}, \ldots, e_{m+p}, e\p_1, \ldots, e\p_{2n}, e\p_{2n+1}, \ldots, e\p_{2n+2q}$ generate the super Clifford algebra $\mathcal{C}_{m+p,2n+2q}$. The superdimensions of the vector variables {\bf x} and {\bf y} are denoted by $M=m-2n$ and $P:=p-2q$ respectively. Thus the supervector ${\bf x}+{\bf y}$ has superdimension $M+P$ and satisfy that $({\bf x}+{\bf y})^2={\bf x}^2+{\bf y}^2$ since {\bf x} and {\bf y} are {\it orthogonal}, i.e. ${\bf x}{\bf y}=-{\bf y}{\bf x}$.

In this section we {obtain} a suitable Taylor expansion for monogenic superfunctions of the two vector variables ${\bf x}$ and ${\bf y}$, so that ${\bf y}$ is considered as a parameter and that we have a Taylor decomposition in the variable ${\bf x}$. To that end, we study the conditions under which a set of analytic superfunctions $\{F_j({\bf y})\}_{j\in\N_0}$ in an open domain $\Om_p\inc \R^p$ allows for the convergence of the series $\sum_{j=0}^\infty {\bf x}^j F_j({\bf y})$ in a neighborhood $\Om\inc \R^{m+p}$ of $\Om_p$, in such a way that this sum is a monogenic function. As we shall see, the domain of convergence $\Om$ of this power series is 
a SO$(m)$-invariant $(m+p)$-dimensional neighborhood of $\Om_p$, such that the intersections of $\Om$ with all subspaces parallel to $\R^m$ are convex. We refer to this kind of domain as an SO$(m)${\it -normal neighborhood of $\Om_p$}. All {this matter} is summarized in the following 
generalized Cauchy-Kovalevskaya property in the superspace framework.

\begin{teo}\label{Gen_CK_extension}
Let $\{F_j({\bf y})\}_{j\in\N_0}$ be a set of analytic (Clifford-valued) superfunctions in an open domain $\Om_p\inc \R^p$ such that the series
 \begin{equation}\label{HSP}
 F({\bf x},{\bf y})=\sum_{j=0}^\infty {\bf x}^j F_j({\bf y}),
 \end{equation}
 is convergent in an $(m+p)$-dimensional neighborhood of $\Om_p$ and its sum is monogenic,  i.e.\ $(\pa_{\bf x}+ \pa_{\bf y})F=0$. Then the sum  $F({\bf x},{\bf y})$ can be formally written as follows.
\begin{itemize}
\item[$i)$] If $M\notin -2\N_0$ then
\[F({\bf x},{\bf y})= \Gamma\left(\frac{M}{2}\right) \left(\frac{|{\bf x}| \sqrt{\Del_{\bf y}}}{2}\right)^{-\frac{M}{2}}  \hspace{-.1cm}\left(  \frac{|{\bf x}| \sqrt{\Del_{\bf y}}}{2}   J_{\frac{M}{2}-1}\left(|{\bf x}| \sqrt{\Del_{\bf y}}\right)- \frac{{\bf x} \pa_{\bf y}}{2} J_{\frac{M}{2}}\left(|{\bf x}| \sqrt{\Del_{\bf y}}\right) \right) F_0({\bf y}). \]
\item[$ii)$] If $M=-2k\in -2\N_0$ and $m\neq 0$, then $\pa_{\bf y}^{2k+1} F_0 =0$ and 
\begin{multline*}
F({\bf x},{\bf y}) = \frac{1}{k!} \left( P_k\left( \frac{|{\bf x}|^2 \Del_{\bf y}}{4} \right) + \frac{{\bf x} \pa_{\bf y}}{2}  P_{k-1}\left( \frac{|{\bf x}|^2 \Del_{\bf y}}{4} \right) \right) F_0({\bf y})  \\
+ k! \, {\bf x}^{2k+1} \left(\frac{|{\bf x}| \sqrt{\Del_{\bf y}}}{2}\right)^{-k-1}  \hspace{-.1cm}\left(  \frac{|{\bf x}| \sqrt{\Del_{\bf y}}}{2}   J_{k}\left(|{\bf x}| \sqrt{\Del_{\bf y}}\right)+ \frac{{\bf x} \pa_{\bf y}}{2} J_{k+1}\left(|{\bf x}| \sqrt{\Del_{\bf y}}\right) \right) F_{2k+1}({\bf y}).
\end{multline*}
In the case $k=0$, i.e.\ $m=2n$, {we take $P_{-1}\equiv 0$ by convention.}
\item[$iii)$] If $M=-2n\in -2\N_0$, i.e.\ $m=0$, then $\pa_{\bf y}^{2n+1} F_0 =0$ and 
\[
F(\m{x}\p,{\bf y}) = \frac{1}{n!} \left( P_n\left( \frac{-\m{x}\p^{\,2} \Del_{\bf y}}{4} \right) + \frac{\m{x}\p \, \pa_{\bf y}}{2}  P_{n-1}\left( \frac{-\m{x}\p^{\,2} \Del_{\bf y}}{4} \right) \right) F_0({\bf y}).
\]
\end{itemize}
Moreover, the domain of convergence of the of the series (\ref{HSP}) is an SO$(m)$-normal neighborhood $\Om_p$. The symbol $\sqrt{\Del_{\bf y}}$ denotes the square root of the Laplacian $\Del_{\bf y}$ with respect to the vector variable ${\bf y}$ (of which only even powers occur in the above expressions).

\end{teo}

\begin{remark}
The formula provided in $i)$ is an extension of the generalized CK-extension formula (\ref{CKTrad}) in the purely bosonic case  $M=m$ (see also \cite{MR1169463}) to the more general case $M\notin -2\N_0$. 

\noindent In the second statement $ii)$, $F({\bf x},{\bf y})$ does not longer depend on one but on two initial functions. 
{As we shall see in the following proof, if $M\in-2\N_0$, there exists a power ${\bf x}^j$ ($j\in\N$) in the kernel of $\pa_{\bf x}$. This is an important difference with the purely bosonic case, and it allows for an splitting of the series  (\ref{HSP}) into two independent monogenic series.}
\end{remark}

\pf 
Let us apply first the operator $\pa_{\bf x}+\pa_{\bf y}$ to the (formal) sum $F({\bf x}, {\bf y})$. Using formula (\ref{ActDirOnPow}) we obtain
\begin{align}\label{FundCondMonSer}
\left(\pa_{\bf x}+\pa_{\bf y}\right) F({\bf x}, {\bf y}) \nonumber
&= \sum_{j=0}^\infty \pa_{\bf x}[ {\bf x}^j] \, F_j({\bf y}) + (-1)^j {\bf x}^j\, \pa_{\bf y} [F_j]({\bf y}) \\ \nonumber
&= \sum_{j=1}^\infty c(M,j){\bf x}^{j-1} \, F_j({\bf y}) + \sum_{j=0}^\infty (-1)^j {\bf x}^j\, \pa_{\bf y} [F_j]({\bf y}) \\ 
&= \sum_{j=0}^\infty {\bf x}^j \Big( c(M,j+1) F_{j+1}({\bf y})+ (-1)^j \, \pa_{\bf y} [F_j]({\bf y}) \Big).
 \end{align}
Now we {divide} the proof into three different cases.

\paragraph{Case $i)$ $M\notin -2\N_0$.}  The super-polynomials ${\bf x}^j$ are non-zero and linearly independent. Thus from the condition  $(\pa_{\bf x}+ \pa_{\bf y})F=0$ and (\ref{FundCondMonSer}) we obtain the recurrence relations 
\begin{equation}\label{15}
F_{j+1}({\bf y})= \frac{(-1)^{j+1}}{c(M,j+1) } \, \pa_{\bf y} [F_j]({\bf y}), \;\;\;\;\;\; j\in \N_0.
\end{equation}
Since $M\notin -2\N_0$, it is clear that $c(M,j+1) \neq 0$ for every $j\in \N_0$. The functions $F_j({\bf y})$ are hence uniquely determined by the formulas
\begin{equation}\label{ExF}
F_{2j}({\bf y}) = \frac{(-1)^j \,  \Gam\left(\frac{M}{2}\right)}{2^{2j} j! \Gam\left(\frac{M}{2}+j \right) } \pa^{2j}_{\bf y}[F_0]({\bf y}), \;\;\;\; \mbox{ and } \;\;\;\; F_{2j+1}({\bf y}) = \frac{(-1)^{j+1} \,  \Gam\left(\frac{M}{2}\right)}{2^{2j+1} \, j! \, \Gam\left(\frac{M}{2}+j+1 \right) } \pa^{2j+1}_{\bf y}[F_0]({\bf y}). 
\end{equation}
Substituting (\ref{ExF}) into (\ref{HSP}) we get
\[
F({\bf x}, {\bf y}) = \Gam\left(\frac{M}{2}\right) \left(\sum_{j=0}^\infty \frac{(-1)^j \, \left(|{\bf x}| \sqrt{\Del_{\bf y}}\right)^{2j}}{2^{2j} j! \, \Gam\left(\frac{M}{2}+j \right) } - \frac{{\bf x} \pa_{\bf y}}{2} \sum_{j=0}^\infty \frac{(-1)^j \, \left(|{\bf x}| \sqrt{\Del_{\bf y}}\right)^{2j}}{ 2^{2j} j! \, \Gam\left(\frac{M}{2}+j +1 \right) } \right) [F_0]({\bf y}).
\] 
Using the expansion $ \left(\frac{z}{2}\right)^{-\nu} J_\nu(z)= \sum_{j=0}^\infty \frac{(-1)^j \, z^{2j}}{2^{2j} j! \, \Gam\left(\nu+j +1 \right)}$ we obtain the formula in the stament $i)$.

\paragraph{Case $ii)$ $M=-2k$ and $m\neq 0$.} In this case, the supervector ${\bf x}$ is still non-nilpotent and all the powers ${\bf x}^j$ are linearly independent super-polynomials. Therefore, similarly to (\ref{15}), we obtain that $(\pa_{\bf x}+ \pa_{\bf y})F=0$ implies  
\[c(-2k,j+1) \, F_{j+1}({\bf y})= (-1)^{j+1} \, \pa_{\bf y} [F_j]({\bf y}), \;\;\;\;\;\; j\in \N_0.\]
However, this recurrence formula has an important difference with the previous case. For $j=2k$, we have that $c(-2k,2k+1)=0$, see (\ref{ActDirOnPow}). Hence the above condition reads as
\begin{align*}
F_{j+1}({\bf y}) &= \frac{(-1)^{j+1}}{c(-2k,j+1) } \, \pa_{\bf y} [F_j]({\bf y}), & 0&\leq j\leq 2k-1,\\
 \pa_{\bf y} [F_{2k}]({\bf y}) &=0, \\
 F_{2k+j+2}({\bf y}) &= \frac{(-1)^{j}}{c(-2k,2k+j+2) } \, \pa_{\bf y} [F_{2k+j+1}]({\bf y}), & j&\in\N_0.
\end{align*}
The functions $F_j({\bf y})$ can then be explicitly computed as
\begin{align*}
F_{2j}({\bf y})  &= \frac{(k-j)!}{2^{2j} j! k!} \, \pa_{\bf y}^{2j} [F_0]({\bf y}), & 1&\leq j\leq k, \\
F_{2j+1}({\bf y})  &= \frac{(k-j-1)!}{2^{2j+1} j! k!} \, \pa_{\bf y}^{2j+1} [F_0]({\bf y}), & 1&\leq j\leq k-1,
\end{align*}
and
\begin{align*}
F_{2k+2j+1}({\bf y})  &= \frac{(-1)^j \, k!}{2^{2j} j! (k+j)!} \, \pa_{\bf y}^{2j} [F_{2k+1}]({\bf y}), & j&\in\N_0, \\
F_{2k+2j+2}({\bf y})  &=  \frac{(-1)^j \, k!}{2^{2j+1} j! (k+j+1)!} \, \pa_{\bf y}^{2j+1} [F_{2k+1}]({\bf y}), & j&\in\N_0.
\end{align*}
This means that, in this case, the monogenic sum $F({\bf x},{\bf y})$ depends of two initial functions namely $F_0$ (which must satisfy $\pa_{\bf y}^{2k+1} F_0 =0$) and $F_{2k+1}$. Substituting the above formulas into (\ref{HSP}) we obtain
\begin{align*}
F({\bf x},{\bf y}) &= \frac{1}{k!} \left(\sum_{j=0}^k \frac{(k-j)!}{j!} \left( \frac{|{\bf x}|^2 \Del_{\bf y}}{4} \right)^j  + \frac{{\bf x} \pa_{\bf y}}{2} \sum_{j=0}^{k-1} \frac{(k-1-j)!}{j!} \left( \frac{|{\bf x}|^2 \Del_{\bf y}}{4} \right)^j  \right) [F_0]({\bf y}) \\
&\phantom{=} + k! \, {\bf x}^{2k+1}  \left(\sum_{j=0}^\infty \frac{(-1)^j \, \left(|{\bf x}| \sqrt{\Del_{\bf y}}\right)^{2j}}{2^{2j} j! \, \Gam\left(k+j +1\right) }  + \frac{{\bf x} \pa_{\bf y}}{2} \sum_{j=0}^\infty \frac{(-1)^j \, \left(|{\bf x}| \sqrt{\Del_{\bf y}}\right)^{2j}}{2^{2j} j! \, \Gam\left(k+j +2\right) }  \right) [F_{2k+1}]({\bf y}),
\end{align*}
which yields the formula in statement $ii)$.

\paragraph{Case $iii)$ $M=-2n$, i.e.\ $m=0$.} In this case the vector ${\bf x}=\m{x}\p$ is nilpotent, i.e.\ $\m{x}\p^{\,2n+1}=0$. Thus, the sum in (\ref{HSP}) becomes $F({\bf x},{\bf y})=\sum_{j=0}^{2n} {\bf x}^j F_j({\bf y})$. We then obtain from (\ref{FundCondMonSer}) that $(\pa_{\m{x}\p} +\pa_{\bf y}) F=0$ implies
\begin{align*}
c(-2n,j+1) \, F_{j+1}({\bf y})&= (-1)^{j+1} \, \pa_{\bf y} [F_j]({\bf y}), & 0&\leq j\leq 2n-1,\\[+.1cm] 
\pa_{\bf y}[F_{2n}]({\bf y})&=0.
\end{align*}
Following a similar reasoning as in $ii)$, we get that $F({\bf x}, {\bf y})$ depends only on the initial function $F_0({\bf y})$, which must satisfy $\pa_{\bf y}^{2n+1} F_0 =0$, and
\begin{align*}
F({\bf x},{\bf y}) &= \frac{1}{n!} \left(\sum_{j=0}^n \frac{(n-j)!}{j!} \left( \frac{-\m{x}\p^{\,2} \Del_{\bf y}}{4} \right)^j  + \frac{\m{x}\p \pa_{\bf y}}{2} \sum_{j=0}^{n-1} \frac{(n-1-j)!}{j!} \left( \frac{-\m{x}\p^{\,2} \Del_{\bf y}}{4} \right)^j  \right) [F_0]({\bf y}),
\end{align*}
which is exactly the formula in the statement $iii)$.

Finally, we have to show that the series obtained in $i)$ and $ii)$ actually converge (in the sense of (\ref{series})-(\ref{SerClifSF})) in a neighborhood of $\Om_p$ {in $\R^{m+p}$}. To this end, it is enough to show the convergence of the series 
\begin{equation}\label{SerAna}
\sum_{j=0}^\infty \frac{(-1)^j \, |{\bf x}|^{2j} \Del_{\bf y}^{j}}{2^{2j} j! \, \Gam\left(\frac{M}{2}+j \right) } [F_0]({\bf y}), \;\;\;\; \mbox{with } \;\; F_0\in \mathcal{A}(\Om_p)\otimes \mathfrak{G}_{2q}\otimes \mathcal{C}_{p,2q},
\end{equation}
being the other cases analogous. 

\noindent The Clifford-valued function $ F_0 ({\bf y})$ can be written as $F_0 ({\bf y})=  \sum_{r=1}^s F_{r,0}( { \bf y})\,  a_r$, where $\{a_1,\ldots, a_s\}\inc B$ is a finite set of basis elements of $\mathcal{C}_{p,2q}$, and $F_{r,0}( { \bf y}) \in \mathcal{A}(\Om_p)\otimes \mathfrak{G}_{2n}$ for all $r=1,\ldots,s$. 
To prove the convergence of the series (\ref{SerAna}), it thus suffices to prove the convergence of a series of the form {(see (\ref{SerClifSF})) }
\begin{equation*}
S:=\sum_{j=0}^\infty \frac{(-1)^j \, |{\bf x}|^{2j} \Del_{\bf y}^{j}}{2^{2j} j! \, \Gam\left(\frac{M}{2}+j \right) } [F]({\bf y}), \;\;\;\; \mbox{with } \;\; F\in \mathcal{A}(\Om_p)\otimes \mathfrak{G}_{2q}.
\end{equation*}
Recall that $(-1)^j |{\bf x}|^{2j}=\left(\m{x}^2+\m{x}\p^{\,2}\right)^j$. Therefore
\begin{align*}
S&= \sum_{j=0}^\infty  \sum_{\el=0}^{\min(j,n)} \binom{j}{\el} \frac{\m{x}^{2j-2\el} \m{x}\p^{\, 2\el}}{2^{2j} j! \, \Gam\left(\frac{M}{2}+j \right) } \Del_{\bf y}^{j}[F]({\bf y}) \\
&= \sum_{\el=0}^n\sum_{j=\el}^\infty  \frac{\m{x}^{2j-2\el}}{(j-\el)!} \frac{\m{x}\p^{\, 2\el}}{ \el!} \frac{\Del_{\bf y}^{j}}{2^{2j} \, \Gam\left(\frac{M}{2}+j \right)} [F]({\bf y}) \\
&= \sum_{\el=0}^n \frac{\m{x}\p^{\, 2\el}}{2^{2\el} \el!} \left(\sum_{j=0}^\infty  \frac{\m{x}^{2j}}{j!}  \frac{\Del_{\bf y}^{j+\el}}{2^{2j} \, \Gam\left(\frac{M}{2}+j+\el\right)} [F]({\bf y}) \right).
\end{align*}
To prove the convergence of $S$, it suffices  to prove the convergence of the innermost series in the above expression that we will denote by $S_\el$ for $\el=0,\ldots, n$. {Recall} that $ \Del_{\bf y}= \Del_{\m{y}} + \Del_{\m{y}\p}$ where $ \Del_{\m{y}}= \sum_{j=0}^p\pa_{y_j}^2$ and $\Del_{\m{y}\p}=-4\sum_{j=1}^q \pa_{y\p_{2j-1}}\pa_{y\p_{2j}}$
are the bosonic and fermionic Laplacians with respect to ${\bf y}$ respectively. Then,
\begin{align*}
S_\el&:= \sum_{j=0}^\infty  \frac{\m{x}^{2j}}{j!}  \frac{\Del_{\bf y}^{j+\el}}{2^{2j} \, \Gam\left(\frac{M}{2}+j+\el\right)} [F]({\bf y})  \\
&= \sum_{j=0}^\infty  \sum_{r=0}^{\min(q,j+\el)}  \frac{\m{x}^{2j}}{j!}  \frac{\binom{j+\el}{r} \, \Del_{\m{y}\p}^{r} \, \Del_{\m{y}}^{j+\el-r}}{2^{2j} \, \Gam\left(\frac{M}{2}+j+\el\right)} [F]({\bf y}) \\
 &=\sum_{r=0}^{q} \frac{\Del_{\m{y}\p}^{r}}{r!} \left(\sum_{j=\max(0,r-\el)}^\infty  \frac{\m{x}^{2j}}{j!}  \frac{(j+\el)!}{(j+\el-r)!}  \frac{\Del_{\m{y}}^{j+\el-r}}{2^{2j} \, \Gam\left(\frac{M}{2}+j+\el\right)} [F]({\bf y})\right).
\end{align*}
Now, the convergence of $S_\el$ follows form the convergence of the innermost series in the above expression. According to (\ref{series}), 
{it is enough to prove the convergence of the same series of operators acting on each of the purely bosonic components $F_A(\m{y})$ of $F({\bf y})$. To this end, it suffices to note that $F_A(\m{y})$ is analytic on $\Om_p$ if and only if for every compact $K\inc \Om$ there exists constants $C_K, \lan_K>0$ such that }
%
%
\[
\left| \Del_{\m{y}}^{j+\el-r} F_A(\m{y})\right| \leq C_K\, (2j+2\el-2r)! \,\lan_K^{2j+2\el-2r}, \hspace{.5cm} \mbox{for all } \m{y}\in K, \,A\inc\{1,\ldots, 2q\}.
\]
Hence, for $\m{y}\in K$ we have 
\begin{multline*}
\sum_{j=\max(0,r-\el)}^\infty \left| \frac{\m{x}^{2j}}{j!}  \frac{(j+\el)!}{(j+\el-r)!}  \frac{\Del_{\m{y}}^{j+\el-r}}{2^{2j} \, \Gam\left(\frac{M}{2}+j+\el\right)} [F_A](\m{y})\right| \\
 \leq C_K \lan_K^{2\el-2r} \sum_{j=\max(0,r-\el)}^\infty \frac{(|\m{x}| \lan_K)^{2j}\, (j+\el)! (2j+2\el-2r)!}{j! \, (j+\el-r)! \, 2^{2j} \, \Gam\left(\frac{M}{2}+j+\el\right)}.
\end{multline*}
The series in the right-hand side of the above inequality converges (by the ratio test) if $|\m{x}|  <1/\lan_K$. Therefore the series $S_\el$, S, and finally (\ref{SerAna}) converge normally in a SO$(m)$-normal neighborhood of $\Om_p$. $\hfill\square$

{After minor modifications,} Theorem \ref{Gen_CK_extension} provides a CK-extension principle for another notion of monogenicity associated to the generalized Cauchy-Riemann operator $\pa_{\bf x}-\pa_{x_0}$, where $x_0$ is an extra commuting variable. In this case, monogenicity is still a refinement of harmonicity since  $\pa_{\bf x}-\pa_{x_0}$ factorizes the super Laplace operator in superdimension $M+1$
\begin{equation}\label{LapM+1}
(\pa_{\bf x}-\pa_{x_0})(-\pa_{\bf x}-\pa_{x_0})=\pa_{x_0}^2+\Del_{m|2n}=\Del_{m+1|2n}.
\end{equation}
We obtain the following corollary of Theorem \ref{Gen_CK_extension} when the functions ${F_j}$ are analytic real-valued functions of the real variable $x_0$.


\begin{cor}\label{CK-ext_x_0}
Let $\{F_j(x_0)\}_{j\in\N_0}$ be a set of analytic real-valued functions on a real domain $\Om_1\inc\R$, such that the series $F(x_0,{\bf x})= \sum_{j=0}^\infty {\bf x}^j F_j(x_0)$ converges in a $(m+1)$-dimensional neighborhood of $\Om_1$ {and its sum is monogenic,  i.e.\ $(\pa_{\bf x}-\pa_{x_0})F=0$. Then the sum $F(x_0,{\bf x})$ can be written as follows.}
\begin{itemize}
\item[$i)$] If $M\notin -2\N_0$ then
\[F(x_0,{\bf x})= \Gamma\left(\frac{M}{2}\right) \left(\frac{|{\bf x}| \pa_{x_0}}{2}\right)^{-\frac{M}{2}}  \hspace{-.1cm}\left(  \frac{|{\bf x}| \pa_{x_0}}{2}   J_{\frac{M}{2}-1}\left(|{\bf x}| \pa_{x_0}\right) + \frac{{\bf x} \pa_{x_0}}{2} J_{\frac{M}{2}}\left(|{\bf x}| \pa_{x_0}\right) \right) F_0(x_0). \]
\item[$ii)$] If $M=-2k\in -2\N_0$ and $m\neq 0$, then $\pa_{x_0}^{2k+1} F_0 =0$ and moreover
\begin{multline*}
F(x_0,{\bf x}) = \frac{1}{k!} \left( P_k\left( \frac{|{\bf x}|^2 \pa_{x_0}^2}{4} \right) - \frac{{\bf x} \pa_{x_0}}{2}  P_{k-1}\left( \frac{|{\bf x}|^2 \pa_{x_0}^2}{4} \right) \right) F_0(x_0)  \\
+ k! \, {\bf x}^{2k+1} \left(\frac{|{\bf x}| \pa_{x_0}}{2}\right)^{-k-1}  \hspace{-.1cm}\left(  \frac{|{\bf x}| \pa_{x_0}}{2}   J_{k}\left(|{\bf x}| \pa_{x_0}\right) + \frac{{\bf x} \pa_{x_0}}{2} J_{k+1}\left(|{\bf x}| \pa_{x_0}\right) \right) F_{2k+1}(x_0).
\end{multline*}
\item[$iii)$] If $M=-2n\in -2\N_0$, i.e.\ $m=0$, then $\pa_{x_0}^{2n+1} F_0 =0$ and moreover
\[F(x_0,{\bf x}) = \frac{1}{n!} \left( P_n\left( \frac{-\m{x}\p^{\,2} \pa_{x_0}^2}{4} \right) - \frac{\m{x}\p\,\pa_{x_0}}{2}  P_{n-1}\left( \frac{-\m{x}\p^{\,2} \pa_{x_0}^2}{4} \right) \right) F_0(x_0).\]
\end{itemize}
\end{cor}

\pf Let us consider a new orthogonal Clifford element $e_0$ that together with  $e_1,\ldots,e_m,e\p_1,\ldots,e\p_{2n}$ generate the super Clifford algebra $\mathcal{C}_{m+1, 2n}$. Then the condition $(\pa_{\bf x}-\pa_{x_0})F=0$ is equivalent to $(\pa_{\bf x} e_0-\pa_{x_0}e_0) (e_0F)=0$. The result follows from applying Theorem \ref{Gen_CK_extension} to the function $e_0 F$ and making the identifications ${\bf x}\mapsto {\bf x}e_0$, ${\bf y}\mapsto x_0e_0$  and $\pa_{\bf y}\mapsto -e_0 \pa_{x_0}$.
$\hfill\square$


\section{Plane wave decomposition of the generalized CK-extension}\label{PWDSect}
The generalized CK-extensions formulas studied in the previous section are very related to integration over the supersphere. Indeed, the formulas in Theorem \ref{Gen_CK_extension} are given in terms of the Bessel functions $J_\nu$ and the  Appell polynomials $P_k$. This pattern also appears in the Pizzetti formulas (\ref{PizzSS}) and (\ref{NormInt-2k}) for the integral and normalized integral of polynomials over the supersphere. In this section, we 
{express} the generalized CK-extension in superspace in terms of (normalized) integrals { over the supersphere of functions of plane wave type}. The latter are functions depending on the inner product $\langle{\bf x}, \bf {w} \rangle$, {where the supervector variable ${\bf w}=\m{w}+\m{w}\p=\sum_{j=1}^m w_je_j+\sum_{j=0}^{2n}w\p_j e\p_j$ is independent of ${\bf x}$.}

\begin{teo}{\bf [Plane wave decomposition]}\label{PWDecompCK}
Let $F({\bf x},{\bf y})=\sum_{j=0}^\infty {\bf x}^j F_j({\bf y})$ satisfy the hypotheses of Theorem \ref{Gen_CK_extension}. Then $F({\bf x},{\bf y})$ can be decomposed into plane waves as follows.
\begin{itemize}
\item[$i)$] If $M\notin -2\N_0$ then
\[F({\bf x},{\bf y})= \left(\frac{1}{\sigma_M} \int_{\Sa^{m-1,2n}} \exp\left(-\langle {\bf w}, {\bf x}\rangle \, {\bf w}  \, \pa_{\bf y}\right) \, dS_{\bf w} \right) F_0({\bf y}). \]
\item[$ii)$] If $M=-2k\in -2\N_0$ and $m\neq 0$, then $\pa_{\bf y}^{2k+1} F_0 =0$ and 
\begin{multline*}
F({\bf x},{\bf y})= \left(\frac{1}{\sigma_{-2k}} \int_{\Sa^{m-1,2n}} \exp\left(-\langle {\bf w}, {\bf x}\rangle \, {\bf w}  \, \pa_{\bf y}\right) \, dS_{\bf w} \right) F_0({\bf y})\\
 + (-1)^{k+1} k! (4\pi)^k \left( \int_{\Sa^{m-1,2n}} \exp\left(-\langle {\bf w}, {\bf x}\rangle \, {\bf w}  \, \pa_{\bf y}\right) \, dS_{\bf w} \right) A({\bf y}),
\end{multline*}
where the superfunction $A({\bf y}) \in \mathcal{A}(\Om_p)\otimes \mathfrak{G}_{2q} \otimes \mathcal{C}_{p,2q}$ is such that $\pa_{\bf y}^{2k+1} A = F_{2k+1}$.
\item[$iii)$] If $M=-2n\in -2\N_0$, i.e.\ $m=0$, then $\pa_{\bf y}^{2n+1} F_0 =0$ and 
\[
F(\m{x}\p,{\bf y}) =  \left(\frac{1}{\sigma_{-2n}} \int_{\Sa^{-1,2n}} \big[ \cosh\left(\langle \m{w}\p, \m{x}\p \rangle \, \pa_{\bf y} \right) - \m{w}\p \sinh\left(\langle \m{w}\p, \m{x}\p \rangle \, \pa_{\bf y} \right) \big] \, dS_{\bf w} \right) F_0({\bf y}).\]
\end{itemize}
\end{teo}
\begin{remark}\label{rem5.1}
The plane wave used in $iii)$ is not very different from the one used in $i)$ and $ii)$. {Indeed, for any pair $\m{\xi}, \m{\nu}$ of orthogonal vectors  in $\R^m$ (i.e. $\m{\xi} \m{\nu}= - \m{\nu}\m{\xi}$) with $\m{\xi}^2=-1$ one has the identity 
\[\exp(\m{\xi} \m{\nu}) = \cosh( \m{\nu}) + \m{\xi} \sinh(\m{\nu}).\]
If $m\neq 0$, one can formally replace the role of the unit vector $\m{\xi}\in \Sa^{m-1}$ by $-{\bf w}/|{\bf w}|$, and the role of $\m{\nu}$ by $\langle {\bf w}, {\bf x}\rangle  \, \pa_{\bf y}$, to obtain}
\[\int_{\Sa^{m-1,2n}} \exp\left(-\langle {\bf w}, {\bf x}\rangle \, {\bf w}  \, \pa_{\bf y}\right) \, dS_{\bf w} =  \int_{\Sa^{m-1,2n}} \big[ \cosh\left(\langle{\bf w}, {\bf x} \rangle \, \pa_{\bf y} \right) - {\bf w} \sinh\left({\bf w},{\bf x} \rangle \, \pa_{\bf y} \right) \big] \, dS_{\bf w}.\]
This doesn't hold if $m=0$ since in that case ${\bf w}= \m{w}\p$ is nilpotent and therefore, the correspondence $\m{\xi} \mapsto {\bf w}/|{\bf w}|$ cannot be considered. In particular, the normalized integral of  $\exp\left(-\langle  \m{w}\p,  \m{x}\p\rangle \,  \m{w}\p  \, \pa_{\bf y}\right)F_0({\bf y})$ yields a polynomial of degree $n$ in $\m{x}\p$, while the normalized integral of $[ \cosh\left(\langle \m{w}\p, \m{x}\p \rangle \, \pa_{\bf y} \right) - \m{w}\p \sinh\left(\langle \m{w}\p, \m{x}\p \rangle \, \pa_{\bf y} \right) ]F_0({\bf y})$ is a polynomial of degree $2n$ in $\m{x}\p$. 
\end{remark}

\pf
Let us consider first
\begin{equation}\label{ExpOp}
\exp\left(-\langle {\bf w}, {\bf x}\rangle \, {\bf w}  \, \pa_{\bf y}\right)[F_0]({\bf y}) = \sum_{j=0}^\infty (-1)^j \frac{\langle {\bf w}, {\bf x}\rangle^j}{j!} ({\bf w}  \, \pa_{\bf y})^j [F_0]({\bf y})
\end{equation}
as a power series on the variables $w_1, \ldots, w_m, w\p_1,\ldots, w\p_{2n}$, where ${\bf x}$ and ${\bf y}$ are considered as parameters. 
{If  $m=0$,} it is clear that the above series reduces to a finite sum. If $m\neq 0$, we can show, using a similar reasoning as in the proof of Theorem \ref{Gen_CK_extension}, that (\ref{ExpOp}) converges normally for $\m{w}\in\Sa^{m-1}$ (together with all its derivatives) provided that $(\m{x},\m{y})$ belongs to a certain SO$(m)$-normal neighborhood of $\Om_p$. This condition allows for integration term by term of the series (\ref{ExpOp}) over the supersphere with respect to ${\bf w}$, see (\ref{IntSupSph}). 
{We denote this integral by $I_M[F_0]({\bf x}, {\bf y})$ and compute it as follows}
\begin{align*}
I_M[F_0]({\bf x}, {\bf y}) &:=\left( \int_{\Sa^{m-1,2n}} \exp\left(-\langle {\bf w}, {\bf x}\rangle \, {\bf w}  \, \pa_{\bf y}\right) \, dS_{\bf w} \right) F_0({\bf y}) \nonumber \\ 
 &= \sum_{j=0}^\infty \; \int_{\Sa^{m-1,2n}} \frac{(-1)^j  \langle {\bf w}, {\bf x}\rangle^j}{j!} \, ({\bf w}  \, \pa_{\bf y})^j [F_0]({\bf y}) \, dS_{\bf w} \nonumber \\
&= \sum_{j=0}^\infty \frac{(-1)^j}{(2j)!} \left(\int_{\Sa^{m-1,2n}}  \langle {\bf w}, {\bf x}\rangle^{2j} \, dS_{\bf w} \right) \Del_{\bf y}^j[F_0]({\bf y}) \nonumber \\
&\phantom{=}  - \sum_{j=0}^\infty \frac{(-1)^j}{(2j+1)!} \left(\int_{\Sa^{m-1,2n}}  \langle {\bf w}, {\bf x}\rangle^{2j+1} {\bf w} \, dS_{\bf w} \right) \pa_{\bf y}\Del_{\bf y}^j[F_0]({\bf y}), \nonumber 
\end{align*}
where we have used Lemma \ref{Int_Lem1} in the last equality. Funk-Hecke Theorem \ref{F-H_The} now yields
\begin{align}\label{Int_M}
I_M[F_0]({\bf x}, {\bf y}) 
&= 2\pi^{\frac{M-1}{2}} \left( \sum_{j=0}^\infty \frac{(-1)^j\, \Gam\left(j+\frac{1}{2}\right)}{(2j)!\, \Gam\left(j+\frac{M}{2}\right)} |{\bf x}|^{2j}\Del_{\bf y}^j - {\bf x}\pa_{\bf y} \sum_{j=0}^\infty \frac{(-1)^j\, \Gam\left(j+\frac{3}{2}\right)}{(2j+1)!\, \Gam\left(j+1+\frac{M}{2}\right)} |{\bf x}|^{2j}\Del_{\bf y}^j\right)[F_0]({\bf y}) \nonumber\\
&= 2\pi^{\frac{M}{2}} \left( \sum_{j=0}^\infty \frac{(-1)^j\,  |{\bf x}|^{2j}\Del_{\bf y}^j}{2^{2j} j!\, \Gam\left(j+\frac{M}{2}\right)} - \frac{{\bf x}\pa_{\bf y}}{2} \sum_{j=0}^\infty \frac{(-1)^j\,  |{\bf x}|^{2j}\Del_{\bf y}^j}{2^{2j} j!\, \Gam\left(j+1+\frac{M}{2}\right)}\right)[F_0]({\bf y}),
\end{align}
{where} we have used the identities $\frac{\Gam\left(j+\frac{1}{2}\right)}{\pi^{1/2} (2j)!}= \frac{1}{2^{2j} j!}$ and $\frac{\Gam\left(j+\frac{3}{2}\right)}{\pi^{1/2} (2j+1)!}= \frac{1}{2^{2j+1} j!}$. {Now we divide the proof into three different cases.}

\paragraph{Case $i)$ $M\notin -2\N_0$.} In this case, it suffices to divide formula (\ref{Int_M}) by $\sigma_M=\frac{2\pi^{M/2}}{\Gam(M/2)}$ to obtain the expression in the statement $i)$ of Theorem \ref{Gen_CK_extension}, which proves the result. 

\paragraph{Case $ii)$ $M=-2k$ and $m\neq 0$.} First observe that $\frac{1}{\sigma_{-2k}} \int_{\Sa^{m-1,2n}} \exp\left(-\langle {\bf w}, {\bf x}\rangle \, {\bf w}  \, \pa_{\bf y}\right) [F_0]({\bf y}) \, dS_{\bf w}$ is well-defined in the sense of (\ref{NormInt-2k})-(\ref{NormIntExt}). Indeed, since $\pa_{\bf y}^{2k+1}F_0=0$, the function $\exp\left(-\langle {\bf w}, {\bf x}\rangle \, {\bf w}  \, \pa_{\bf y}\right) [F_0]({\bf y})$ has the form
\[
\exp\left(-\langle {\bf w}, {\bf x}\rangle \, {\bf w}  \, \pa_{\bf y}\right) [F_0]({\bf y}) = \left(\sum_{j=0}^k \frac{(-1)^j \langle {\bf w}, {\bf x}\rangle^{2j}}{(2j)!} |{\bf w}|^{2j} \Del_{\bf y}^j - {\bf w} \pa_{\bf y} \sum_{j=0}^{k-1} \frac{(-1)^j \langle {\bf w}, {\bf x}\rangle^{2j+1}}{(2j+1)!} |{\bf w}|^{2j} \Del_{\bf y}^j\right)[F_0]({\bf y}),
\]
which is a linear combination of functions of the form $|{\bf w}|^{2j} R$ where $R$ is a homogeneous super-polynomial on ${\bf w}$ of degree at most $2k$.

\noindent {We denote by $NI_{-2k}[F_0]({\bf x}, {\bf y})$ the normalized integral  of this function over the supersphere,  and compute it using the} Funk-Hecke Theorem for normalized integrals (see Corollary \ref{F-H_TheNor}) as follows
\begin{align}\label{ii1}
NI_{-2k}[F_0]({\bf x}, {\bf y}) &:= \frac{1}{\sigma_{-2k}} \int_{\Sa^{m-1,2n}} \exp\left(-\langle {\bf w}, {\bf x}\rangle \, {\bf w}  \, \pa_{\bf y}\right) [F_0]({\bf y}) \, dS_{\bf w} \nonumber \\
&= \sum_{j=0}^k \frac{(-1)^j}{(2j)!} \left( \frac{1}{\sigma_{-2k}} \int_{\Sa^{m-1,2n}} \langle {\bf w}, {\bf x}\rangle^{2j} \, dS_{\bf w}\right) \Del_{\bf y}^j[F_0]({\bf y}) \nonumber \\
&\phantom{=}- \sum_{j=0}^{k-1} \frac{(-1)^j}{(2j+1)!} \left( \frac{1}{\sigma_{-2k}} \int_{\Sa^{m-1,2n}} \langle {\bf w}, {\bf x}\rangle^{2j+1} {\bf w}\, dS_{\bf w}\right) \pa_{\bf y}\Del_{\bf y}^j[F_0]({\bf y}) \nonumber \\
&= \frac{1}{k!} \left( \sum_{j=0}^k \frac{(k-j)!}{2^{2j} j!} |{\bf x}|^{2j} \Del_{\bf y}^j + \frac{{\bf x}\pa_{\bf y}}{2} \sum_{j=0}^{k-1} \frac{(k-j-1)!}{2^{2j} j!} |{\bf x}|^{2j} \Del_{\bf y}^j \right) [F_0]({\bf y}).
\end{align}


\noindent We now proceed to compute the second integral in the right-hand side of {the equality in the statement $ii)$}. From (\ref{Int_M}) we get
\begin{align*}
I_{-2k}[A]({\bf x},{\bf y}) &= 2\pi^{-k} \left( \sum_{j=0}^\infty \frac{(-1)^j\,  |{\bf x}|^{2j}\Del_{\bf y}^j}{2^{2j} j!\, \Gam\left(j-k\right)} - \frac{{\bf x}\pa_{\bf y}}{2} \sum_{j=0}^\infty \frac{(-1)^j\,  |{\bf x}|^{2j}\Del_{\bf y}^j}{2^{2j} j!\, \Gam\left(j+1-k\right)}\right)[A]({\bf y}) \nonumber \\
&= 2\pi^{-k} \left( \sum_{j=k+1}^\infty \frac{(-1)^j\,  |{\bf x}|^{2j}\Del_{\bf y}^j}{2^{2j} j!\, \Gam\left(j-k\right)} - \frac{{\bf x}\pa_{\bf y}}{2} \sum_{j=k}^\infty \frac{(-1)^j\,  |{\bf x}|^{2j}\Del_{\bf y}^j}{2^{2j} j!\, \Gam\left(j+1-k\right)}\right)[A]({\bf y})\nonumber \\
 &=2\pi^{-k} \left( \sum_{j=0}^\infty \frac{(-1)^{j+k+1}\,  |{\bf x}|^{2j+2k+2}\Del_{\bf y}^{j+k+1}}{2^{2j+2k+2} (j+k+1)!\, \Gam\left(j+1\right)} - \frac{{\bf x}\pa_{\bf y}}{2} \sum_{j=0}^\infty \frac{(-1)^{j+k}\,  |{\bf x}|^{2j+2k}\Del_{\bf y}^{j+k}}{2^{2j+2k} (j+k)!\, \Gam\left(j+1\right)}\right)[A]({\bf y})\nonumber \\
 &=\frac{(-1)^{k+1}}{(4\pi)^k} {\bf x}^{2k+1} \left(\sum_{j=0}^\infty \frac{(-1)^{j}\,  |{\bf x}|^{2j}\Del_{\bf y}^{j}}{2^{2j} j!\, \Gam\left(j+k+1\right)} + \frac{{\bf x}\pa_{\bf y}}{2}  \sum_{j=0}^\infty \frac{(-1)^{j}\,  |{\bf x}|^{2j}\Del_{\bf y}^{j}}{2^{2j} j!\, \Gam\left(j+k+2\right)}\right) \pa_{\bf y}^{2k+1}[A]({\bf y}),
\end{align*}
where we have used that $\frac{1}{\Gam(j-k)}=0$ if $j\leq k$. Then we obtain
\begin{multline}\label{I-2k}
(-1)^{k+1}k! (4\pi)^k I_{-2k}[A]({\bf x},{\bf y}) \\= k!{\bf x}^{2k+1} \left(\sum_{j=0}^\infty \frac{(-1)^{j}\,  |{\bf x}|^{2j}\Del_{\bf y}^{j}}{2^{2j} j!\, \Gam\left(j+k+1\right)} + \frac{{\bf x}\pa_{\bf y}}{2}  \sum_{j=0}^\infty \frac{(-1)^{j}\,  |{\bf x}|^{2j}\Del_{\bf y}^{j}}{2^{2j} j!\, \Gam\left(j+k+2\right)}\right)[F_{2k+1}]({\bf y}).
\end{multline}
Combining formulas (\ref{ii1})  and (\ref{I-2k}) we obtain the expression in statement $ii)$ of Theorem \ref{Gen_CK_extension}, which proves the result. 

\paragraph{Case $iii)$ $M=-2n$, i.e.\ $m=0$.} In this case, the plane wave $\exp\left(-\langle \m{w}\p, \m{x}\p\rangle \, \m{w}\p  \, \pa_{\bf y}\right)[F_0]({\bf y})$ is not suitable to describe the generalized CK-extension $F(\m{x}\p,{\bf y})$. Indeed, since $\langle \m{w}\p, \m{x}\p\rangle^j \, \m{w}\p^j =0$ for $j>n$, the {normalized} integral of $\exp\left(-\langle \m{w}\p, \m{x}\p\rangle \, \m{w}\p  \, \pa_{\bf y}\right)[F_0]({\bf y})$ over the supersphere is a polynomial of degree at most $n$ in $\m{x}\p$, while in general $F(\m{x}\p,{\bf y})$ has degree $2n$ in $\m{x}\p$. In order to produce a polynomial of full degree $2n$ we use the plane wave (see Remark \ref{rem5.1})
\[\Big( \cosh\left(\langle{\bf w}, {\bf x} \rangle \, \pa_{\bf y} \right) - {\bf w} \sinh\left({\bf w},{\bf x} \rangle \, \pa_{\bf y} \right) \Big)[F_0]({\bf y}).\]
As in the previous case, 
{we denote by $NI_{-2n}[F_0]({\bf x},{\bf y})$ the normalized integral of the above plane wave, and compute it using the Funk-Hecke theorem,}
\begin{align*}
NI_{-2n}[F_0]({\bf x},{\bf y}) &:= \left(\frac{1}{\sigma_{-2n}} \int_{\Sa^{-1,2n}} \big[ \cosh\left(\langle \m{w}\p, \m{x}\p \rangle \, \pa_{\bf y} \right) - \m{w}\p \sinh\left(\langle \m{w}\p, \m{x}\p \rangle \, \pa_{\bf y} \right) \big] \, dS_{\bf w} \right) F_0({\bf y}) \nonumber \\
&= \frac{1}{\sigma_{-2n}} \int_{\Sa^{-1,2n}} \left(\sum_{j=0}^n \frac{ \langle \m{w}\p, \m{x}\p \rangle^{2j}}{(2j)!} \pa_{\bf y}^{2j} - \sum_{j=0}^{n-1} \frac{ \langle \m{w}\p, \m{x}\p \rangle^{2j+1} \m{w}\p }{(2j+1)!} \pa_{\bf y}^{2j+1}\right) [F_0]({\bf y}) \, dS_{\bf w} \\
&= \sum_{j=0}^n \left( \frac{1}{\sigma_{-2n}} \int_{\Sa^{-1,2n}} \langle \m{w}\p, \m{x}\p \rangle^{2j} \, dS_{\m{w}\p}\right) \frac{\pa_{\bf y}^{2j}}{(2j)!}[F_0]({\bf y}) \\
&\phantom{=} - \sum_{j=0}^{n-1}\left( \frac{1}{\sigma_{-2n}} \int_{\Sa^{-1,2n}} \langle \m{w}\p, \m{x}\p \rangle^{2j+1} \m{w}\p \, dS_{\m{w}\p}\right) \frac{\pa_{\bf y}^{2j+1}}{(2j+1)!} [F_0]({\bf y}) \\
&= \frac{1}{n!} \left( \sum_{j=0}^n \frac{(n-j)!}{2^{2j} j!} (-1)^j {\m{x}\p}^{\,2j} \Del_{\bf y}^j + \frac{\m{x}\p \pa_{\bf y}}{2} \sum_{j=0}^{n-1} \frac{(n-j-1)!}{2^{2j} j!} (-1)^j {\m{x}\p}^{\,2j} \Del_{\bf y}^j \right) [F_0]({\bf y}),
\end{align*}
which proves the result. $\hfill\square$
{The task is now to show} the following plane wave counterpart of Corollary \ref{CK-ext_x_0}.

\begin{cor}\label{PWCKx_0}
Let $F(x_0,{\bf x})= \sum_{j=0}^\infty {\bf x}^j F_j(x_0)$ satisfy the hypotheses of Corollary \ref{CK-ext_x_0}. Then $F(x_0,{\bf x})$ can be decomposed into plane waves as follows.
\begin{itemize}
\item[$i)$] If $M\notin -2\N_0$ then
\[F({x_0}, {\bf x})= \left(\frac{1}{\sigma_M} \int_{\Sa^{m-1,2n}} \exp\left(\langle {\bf w}, {\bf x}\rangle \, {\bf w}  \, \pa_{x_0}\right) \, dS_{\bf w} \right) F_0({x_0}). \]
\item[$ii)$] If $M=-2k\in -2\N_0$ and $m\neq 0$, then $\pa_{x_0}^{2k+1} F_0 =0$ and 
\begin{multline*}
F({x_0}, {\bf x})= \left(\frac{1}{\sigma_{-2k}} \int_{\Sa^{m-1,2n}} \exp\left(\langle {\bf w}, {\bf x}\rangle \, {\bf w}  \, \pa_{x_0}\right) \, dS_{\bf w} \right) F_0(x_0)\\
 + k! (4\pi)^k \left( \int_{\Sa^{m-1,2n}} \exp\left(\langle {\bf w}, {\bf x}\rangle \, {\bf w}  \, \pa_{x_0}\right) \, dS_{\bf w} \right) B({x_0}),
\end{multline*}
where the real-valued function $B({x_0})\in \mathcal{A}(\Om_1)$ is such that $\pa_{x_0}^{2k+1} B = F_{2k+1}$.
\item[$iii)$] If $M=-2n\in -2\N_0$, i.e.\ $m=0$, then $\pa_{x_0}^{2n+1} F_0 =0$ and 
\[
F({x_0}, \m{x}\p) =  \left(\frac{1}{\sigma_{-2n}} \int_{\Sa^{-1,2n}} \big[ \cos\left(\langle \m{w}\p, \m{x}\p \rangle \, \pa_{x_0} \right) + \m{w}\p \, \sin\left(\langle \m{w}\p, \m{x}\p \rangle \, \pa_{x_0} \right) \big] \, dS_{\m{w}\p} \right) F_0({x_0}).\]
\end{itemize}
\end{cor}

\pf Similarly to the proof of Corollary \ref{CK-ext_x_0}, we consider a new orthogonal Clifford element $e_0$ generating, together with  $e_1,\ldots,e_m,e\p_1,\ldots,e\p_{2n}$, the super Clifford algebra $\mathcal{C}_{m+1, 2n}$. The result the follows from  applying Theorem \ref{PWDecompCK} to the function $e_0 F$ and making the identifications ${\bf x}\mapsto {\bf x}e_0$, ${\bf w}\mapsto {\bf w}e_0$, ${\bf y}\mapsto x_0e_0$ and $\pa_{\bf y}\mapsto -e_0 \pa_{x_0}$. The proof of statement $iii)$ makes use of the identities 
\begin{align*}
 \cosh\left(\langle \m{w}\p, \m{x}\p \rangle \, (-e_0\pa_{x_0}) \right) &=  \cos\left(\langle \m{w}\p, \m{x}\p \rangle \,\pa_{x_0} \right), &  \sinh\left(\langle \m{w}\p, \m{x}\p \rangle \, (-e_0\pa_{x_0}) \right) &=  (-e_0) \sin\left(\langle \m{w}\p, \m{x}\p \rangle \,\pa_{x_0} \right).
 \end{align*}

$\hfill\square$

\section{Integration of monogenic plane waves}\label{IMPWSect}
In this section, we study decompositions of certain generalized CK-extensions into plane waves constructed out of holomorphic functions. These ideas will be used in Section \ref{PWDCKSect} where we find a {plane wave decomposition of the super Cauchy kernel.}

Let $g(z)=g_1(a,b)+ig_2(a,b)$ be a holomorphic $\C$-valued function of the complex variable $z=a+ib$ in an open domain $\Om\subseteq \R^2\iso \C$. If $m\neq 0$, given a supervector parameter ${\bf w}=\m{w}+\m{w}\p$, we define 
\begin{equation}\label{HolSF}
g(\langle{\bf x},{\bf w}\rangle - x_0 {\bf w} ) = g_1(\langle{\bf x},{\bf w}\rangle, x_0 |{\bf w}|) - \frac{{\bf w}}{|{\bf w}|} g_2(\langle{\bf x},{\bf w}\rangle, x_0 |{\bf w}|),
\end{equation}
as an element of $C^\infty(\Om_{\m{w}})\otimes \mathfrak{G}_{2n}\otimes \mathcal{C}_{m,2n}$ where $\Om_{\m{w}}=\left\{{(\m{x}, x_0)}\in \R^{m+1}: \left(\langle\m{x},\m{w}\rangle, \, x_0|\m{w}|\right)  \in \Om \right\}$. Here the functions $g_\el(\langle{\bf x},{\bf w}\rangle, x_0 |{\bf w}|)$, $\el=1,2$, are defined as in (\ref{Tay_Ser}).

Since $g_1,g_2$ are real analytic functions in $\Om$, the definition of $g_\el(\langle{\bf x},{\bf w}\rangle, x_0 |{\bf w}|)$, $\el=1,2$, is independent of the splitting of $\langle{\bf x},{\bf w}\rangle$ and $x_0 |{\bf w}|$, {provided that the conditions of Lemma \ref{Lem1} are satisfied}. Therefore, if {$(\underline{0},x_0|\m{w}|) \in \Om$} and $\left(\langle\m{x},\m{w}\rangle, \, x_0|\m{w}|\right)$ belongs to the region of  convergence of the Taylor series of $g_\el$ around $(\underline{0},x_0|\m{w}|)$,   we obtain 
\begin{align}\label{Taylor_PW}
g(\langle{\bf x},{\bf w}\rangle - x_0 {\bf w})&= \sum_{j=0}^\infty \frac{\langle{\bf x},{\bf w}\rangle^j}{j!} \, \pa_a^j g_1 \left(0, x_0 |{\bf w}|\right) - \frac{{\bf w}}{|{\bf w}|} \sum_{j=0}^\infty \frac{\langle{\bf x},{\bf w}\rangle^j}{j!} \, \pa_a^j g_2\left(0, x_0 |{\bf w}|\right) \\
&= \sum_{j=0}^\infty \frac{\langle{\bf x},{\bf w}\rangle^j}{j!} \,  g^{(j)}\left(- x_0 {\bf w}\right),\nonumber
\end{align}
which coincides with the Taylor expansion of $g$ as a function of one complex variable.

{In (\ref{HolSF})},  we have replaced the role of the complex imaginary unit $i$ by {the} supervector $-\frac{{\bf w}}{|{\bf w}|}$. We recall that this correspondence does not exist if $m=0$ since ${\bf w}=\m{w}\p$ is nilpotent. However, it is still possible to find an analogous of definition  (\ref{HolSF}) in this context given by
\begin{equation*}
g(\langle\m{x}\p,\m{w}\p\rangle - x_0 \m{w}\p) = \sum_{j=0}^{2n} \frac{(\langle\m{x}\p,\m{w}\p\rangle - x_0 \m{w}\p)^j}{j!} g^{(j)}(0),
\end{equation*}
for any real-valued function $g$ of class $C^{2n}$ in a neighborhood of $z=0$. Therefore, in the case $m=0$, it suffices to consider only the generators $(\langle\m{x}\p,\m{w}\p\rangle - x_0 \m{w}\p)^j$ for $j=0, 1, \ldots, 2n$.

{A function of the type (\ref{HolSF}) is  called a {\it monogenic plane wave}. The monogenicity of $g(\langle{\bf x},{\bf w}\rangle - x_0 {\bf w} )$ is established in our next Lemma.}

\begin{lem}
The function $g(\langle{\bf x},{\bf w}\rangle - x_0 {\bf w} ) $ is the kernel of the operator $(\pa_{\bf x}-\pa_{x_0})$.
\end{lem}
\pf 
We only prove the result for $m\neq 0$ since the case $m=0$ can be treated analogously. From the chain rule in superspace (see \cite{Berezin:1987:ISA:38130}) we obtain,
\begin{align*}
\pa_{x_0} g(\langle{\bf x},{\bf w}\rangle - x_0 {\bf w} ) &= |{\bf w}| \left(  \pa_b g_1(\langle{\bf x},{\bf w}\rangle, x_0 |{\bf w}|) -  \frac{{\bf w}}{|{\bf w}|} \, \pa_bg_2(\langle{\bf x},{\bf w}\rangle, x_0 |{\bf w}|)\right)\\
\pa_{x_j} g(\langle{\bf x},{\bf w}\rangle - x_0 {\bf w} ) &= w_j \left(  \pa_a g_1(\langle{\bf x},{\bf w}\rangle, x_0 |{\bf w}|) -  \frac{{\bf w}}{|{\bf w}|} \, \pa_ag_2(\langle{\bf x},{\bf w}\rangle, x_0 |{\bf w}|)\right),  &j&=1,\ldots,m, \\
\pa_{x\p_{2j-1}} g(\langle{\bf x},{\bf w}\rangle - x_0 {\bf w} ) &= -\frac{1}{2} w\p_{2j} \left( \pa_a g_1(\langle{\bf x},{\bf w}\rangle, x_0 |{\bf w}|) -  \frac{{\bf w}}{|{\bf w}|} \, \pa_ag_2(\langle{\bf x},{\bf w}\rangle, x_0 |{\bf w}|) \right),   &j&=1,\ldots,n,\\
\pa_{x\p_{2j}} g(\langle{\bf x},{\bf w}\rangle - x_0 {\bf w} ) &= \frac{1}{2} w\p_{2j-1} \left( \pa_a g_1(\langle{\bf x},{\bf w}\rangle, x_0 |{\bf w}|) -  \frac{{\bf w}}{|{\bf w}|} \, \pa_ag_2(\langle{\bf x},{\bf w}\rangle, x_0 |{\bf w}|) \right), &j&=1,\ldots,n. 
\end{align*}
This implies that
\begin{align*}
\pa_{\m{x}} g(\langle{\bf x},{\bf w}\rangle - x_0 {\bf w} ) &= \m{w} \left(  \pa_a g_1(\langle{\bf x},{\bf w}\rangle, x_0 |{\bf w}|) -  \frac{{\bf w}}{|{\bf w}|} \, \pa_ag_2(\langle{\bf x},{\bf w}\rangle, x_0 |{\bf w}|)\right), \\
\pa_{\m{x}\p} g(\langle{\bf x},{\bf w}\rangle - x_0 {\bf w} ) &= -\m{w}\p \left(  \pa_a g_1(\langle{\bf x},{\bf w}\rangle, x_0 |{\bf w}|) -  \frac{{\bf w}}{|{\bf w}|} \, \pa_ag_2(\langle{\bf x},{\bf w}\rangle, x_0 |{\bf w}|)\right),
\end{align*}
yielding
\[
\pa_{{\bf x}} g(\langle{\bf x},{\bf w}\rangle - x_0 {\bf w} ) = \left(\pa_{\m{x}\p} - \pa_{\m{x}}\right)g(\langle{\bf x},{\bf w}\rangle - x_0 {\bf w} ) = -{\bf w}  \, \pa_a g_1(\langle{\bf x},{\bf w}\rangle, x_0 |{\bf w}|) -|{\bf w}| \, \pa_ag_2(\langle{\bf x},{\bf w}\rangle, x_0 |{\bf w}|).
\]
Finally, using the Cauchy-Riemann equations for the functions $g_1,g_2$, we obtain
\begin{align*}
\left(\pa_{\bf x} - \pa_{x_0}\right)g(\langle{\bf x},{\bf w}\rangle - x_0 {\bf w} ) &=-\Big[{\bf w} \left(\pa_a g_1 - \pa_b g_2\right)+ |{\bf w}| \left(\pa_a g_2 + \pa_b g_1\right)\Big] \, \left(\langle{\bf x},{\bf w}\rangle, x_0 |{\bf w}|\right) =0,
\end{align*}
which proves the result. $\hfill\square$

{Now we} proceed  to study monogenic plane wave decompositions of generalized CK-extensions {given by} integrals of the form $\int_{\Sa^{m-1,2n}} g(\langle{\bf x},{\bf w}\rangle - x_0 {\bf w})\, {\bf w}^{\el-1} dS_{\bf w}$, $\el \in \N$. Lemma \ref{Int_Lem1} shows that it suffices to consider these integrals only for $\el=1,2$.

\begin{lem}\label{Int_PW}
Let $m\neq 0$ and let $g$ be an holomorphic function in $\Om\inc\R^2$. Then for $\el=1,2$ one has
\begin{equation}\label{HolPWExp}
\int_{\Sa^{m-1,2n}} g(\langle{\bf x},{\bf w}\rangle - x_0 {\bf w})\, {\bf w}^{\el-1} dS_{\bf w} = \left(\int_{\Sa^{m-1,2n}} \exp\left(\langle {\bf w}, {\bf x}\rangle \, {\bf w}  \, \pa_{b}\right) \, dS_{\bf w} \right) g_{\el}(0,x_0),
\end{equation}
{provided that $(\m{x},x_0)$ belongs to a certain SO$(m)$-normal neighborhood of the intersection $\Om\cap (\{0\}\times \R)$ of $\Om$ with the imaginary axis $\{0\}\times \R$.} 
\end{lem}
\pf Using similar ideas as in the proofs of Theorems \ref{Gen_CK_extension} and \ref{PWDecompCK}, we can show that the series (\ref{Taylor_PW}) converges normally for $\m{w}\in\Sa^{m-1}$ (together with all its derivatives) 
{if} $(\m{x},x_0)$ belongs to certain SO$(m)$-normal neighborhood of the set $\Om\cap (\{0\}\times \R)=\{(0,b): (0,b)\in \Om\}$. Therefore the series expansion 
{of} $g(\langle{\bf x},{\bf w}\rangle - x_0 {\bf w})\, {\bf w}^{\el-1}$ can be integrated term by term over the supersphere with respect to ${\bf w}$. Let us denote this integral by $\mathcal{I}_\el(x_0,{\bf x})$. Using 
Lemma \ref{Int_Lem1} we obtain
\[
\mathcal{I}_\el(x_0,{\bf x}) = \sum_{j=0}^\infty \left(\int_{\Sa^{m-1,2n}} \langle{\bf x},{\bf w}\rangle^j {\bf w}^{\el-1} dS_{\bf w} \right)  \frac{\pa_a^j g_1 \left(0, x_0 \right)}{j!} - \left(\int_{\Sa^{m-1,2n}} \langle{\bf x},{\bf w}\rangle^j {\bf w}^{\el} dS_{\bf w} \right)  \frac{\pa_a^j g_2 \left(0, x_0 \right)}{j!}.
\]
By Funk-Hecke Theorem \ref{F-H_The} we have
\begin{align*}
\mathcal{I}_1(x_0,{\bf x})& =2 \pi^{\frac{M-1}{2}} \sum_{j=0}^\infty \left({\bf x}^{2j} \frac{(-1)^{j}\, \Gam\left(j+\frac{1}{2}\right)}{\Gam\left(j+\frac{M}{2}\right)} \, \frac{\pa_a^{2j} g_1 \left(0, x_0 \right)}{(2j)!}   -    {\bf x}^{2j+1} \frac{(-1)^{j}\, \Gam\left(j+\frac{3}{2}\right)}{\Gam\left(j+1+\frac{M}{2}\right)} \, \frac{\pa_a^{2j+1}g_2\left(0, x_0 \right)}{(2j+1)!}  \right),\\
\mathcal{I}_2(x_0,{\bf x})&= 2 \pi^{\frac{M-1}{2}} \sum_{j=0}^\infty \left( {\bf x}^{2j+1} \frac{(-1)^{j}\, \Gam\left(j+\frac{3}{2}\right)}{\Gam\left(j+1+\frac{M}{2}\right)} \, \frac{\pa_a^{2j+1} g_1 \left(0, x_0 \right)}{(2j+1)!}  +   {\bf x}^{2j} \frac{(-1)^{j}\, \Gam\left(j+\frac{1}{2}\right)}{\Gam\left(j+\frac{M}{2}\right)} \, \frac{\pa_a^{2j} g_2\left(0, x_0 \right)}{(2j)!} \right),
\end{align*}
From the Cauchy-Riemann equations we have the following identities 
\begin{align*}
\pa_a^{2j} g_1 &= (-1)^j \pa_b^{2j} g_1, & \pa_a^{2j+1} g_2 &= -(-1)^j \pa_b^{2j+1} g_1, & \pa_a^{2j+1} g_1 &= (-1)^j \pa_b^{2j+1} g_2, & \pa_a^{2j} g_2 &= (-1)^j \pa_b^{2j} g_2.
\end{align*}
Then,  for $\el=1,2$, we can write
\[
\mathcal{I}_\ell(x_0,{\bf x})=  2 \pi^{\frac{M-1}{2}} \sum_{j=0}^\infty \left({\bf x}^{2j} \frac{ \Gam\left(j+\frac{1}{2}\right)}{\Gam\left(j+\frac{M}{2}\right)} \, \frac{ \pa_b^{2j} }{(2j)!}   + {\bf x}^{2j+1} \frac{ \Gam\left(j+\frac{3}{2}\right)}{\Gam\left(j+1+\frac{M}{2}\right)} \,\frac{\pa_b^{2j+1}}{(2j+1)!}  \right) g_\el(0,x_0).
\]
In this way, we have written the monogenic functions $\mathcal{I}_\ell(x_0,{\bf x})$ as power series in the vector variable ${\bf x}$. Thus, it suffices to apply the plane wave decomposition from Corollary \ref{PWCKx_0} in order to prove formula (\ref{HolPWExp}).
%

\noindent If $M\notin -2\N_0$, the monogenic power series $\mathcal{I}_\ell(x_0,{\bf x})$ is completely determined by the initial function
\[F_0(x_0) = \mathcal{I}_\ell(x_0,0) =   \sigma_M\, g_\el(0,x_0), \;\;\;  \;\;\; \ell=1,2.
\]

\noindent On the other hand, if $M=-2k$, the series $\mathcal{I}_\ell(x_0,{\bf x})$ is determined by the two initial functions
\[F_0(x_0) = \mathcal{I}_\ell(x_0,0) =   \sigma_{-2k}\, g_\el(0,x_0) =0, \;\;\;  \;\;\; \ell=1,2.
\]and 
 \[F_{2k+1}(x_0) = 2 \pi^{-k-\frac{1}{2}}  \frac{\Gam\left(k+\frac{3}{2}\right)}{(2k+1)!}\; \pa_b^{2k+1} g_\el(0,x_0) = \frac{1}{(4\pi)^{k} \, k!} \; \pa_b^{2k+1} g_\el(0,x_0), \;\;\;  \;\;\; \ell=1,2.
\]
In both cases, Corollary \ref{PWCKx_0} yields formula (\ref{HolPWExp}). 
$\hfill\square$

A similar result to Lemma \ref{Int_PW} also holds for the normalized integrals defined in (\ref{NormInt-2k})-(\ref{NormIntExt}) for the cases of even negative superdimensions. 

\begin{lem}\label{NorInt_PW}
{Let $M=-2k$ and $g(z)=z^{2s+\el-1}$ with $s\in \N_0$, $\el=1,2$ and $2s+\el-1\leq 2k$.} Then, if $m\neq 0$, one has
\[\frac{1}{\sigma_{-2k}} \int_{\Sa^{m-1,2n}} {(\langle{\bf x},{\bf w}\rangle - x_0 {\bf w})^{2s+\el-1}}\, {\bf w}^{\el-1} dS_{\bf w} = (-1)^s \left(\frac{1}{\sigma_{-2k}} \int_{\Sa^{m-1,2n}} \exp\left(\langle {\bf w}, {\bf x}\rangle \, {\bf w}  \, {\pa_{x_0}}\right) \, dS_{\bf w} \right) {x_0^{2s+\el-1}}.\]
Similarly, if $m=0$, one has
\begin{multline*}
\frac{1}{\sigma_{-2n}} \int_{\Sa^{-1,2n}} (\langle\m{x}\p,\m{w}\p\rangle - x_0 \m{w}\p)^{2s+\el-1}\, \m{w}\p^{\,\el-1} dS_{\m{w}\p} \\=  (-1)^s \left( \frac{1}{\sigma_{-2n}}\int_{\Sa^{-1,2n}} \big[ \cos\left(\langle \m{w}\p, \m{x}\p \rangle \, \pa_{x_0} \right) + \m{w}\p \sin\left(\langle \m{w}\p, \m{x}\p \rangle \, \pa_{x_0} \right) \big] \, dS_{\m{w}\p} \right) x_0^{2s+\el-1}.
\end{multline*}
%
\end{lem}
\begin{remark}
{Observe that $g(ix_0)=(ix_0)^{2s+\el-1}=(-1)^s x_0^{2s} (ix_0)^{\el-1}$. Therefore, $g_\el(0,x_0)=(-1)^s x_0^{2s+\el-1}$.}
\end{remark}

\pf {
Note that all the normalized integrals in the Lemma are well-defined since they act on polynomials of degree smaller than $2k+1$ on the vector variable ${\bf w}$. The proof of this Lemma follows in much the same way as the proof of Lemma \ref{Int_PW}. Indeed, we may write the polynomial $(\langle{\bf x},{\bf w}\rangle - x_0 {\bf w})^{2s+\el-1}\, {\bf w}^{\el-1}$ as
\[
(\langle{\bf x},{\bf w}\rangle - x_0 {\bf w})^{2s+\el-1}\, {\bf w}^{\el-1} = (2s+\el-1)! \sum_{j=0}^{2s+\el-1} \frac{\langle{\bf x},{\bf w}\rangle^j}{j!} \, \frac{{\bf w}^{2s+2\el-j-2}}{(2s+\el-j-1)!} \, (-x_0)^{2s+\el-j-1}. 
\]}
\noindent Then using Lemma \ref{Int_Lem1} and Corollary \ref{F-H_TheNor} (Funk-Hecke theorem for the normalized integral) we obtain
\begin{multline*}
\frac{1}{\sigma_{-2k}} \int_{\Sa^{m-1,2n}} {(\langle{\bf x},{\bf w}\rangle - x_0 {\bf w})^{2s+\el-1}}\, {\bf w}^{\el-1} dS_{\bf w} \\
=  \left(\sum_{j=0}^{k} {\bf x}^{2j} \frac{(-1)^j (k-j)!}{2^{2j} k! \, j!} \, \pa_{x_0}^{2j} - \sum_{j=0}^{k-1} 
{\bf x}^{2j+1} \frac{(-1)^j (k-j-1)!}{2^{2j+1} k! \, j!} \,\pa_{x_0}^{2j+1} \right)  {\left[(-1)^s x_0^{2s+\el-1}\right]}.
\end{multline*}
Finally, the plane wave decompositions given in Corollary \ref{PWCKx_0} $ii)-iii)$ yield the result.
$\hfill\square$
 
 \begin{remark}\label{RemPW}
 If $m\neq 0$, {Lemma \ref{NorInt_PW} can be reformulated as follows
 \begin{multline*}
 \left(\frac{1}{\sigma_{-2k}} \int_{\Sa^{m-1,2n}} \exp\left(\langle {\bf w}, {\bf x}\rangle \, {\bf w}  \,{\pa_{x_0}}\right) \, dS_{\bf w} \right) {x_0^{2s+\el-1}} \\= (-1)^{\el-1} \frac{(k-s-\el+1)!}{4^{s+\el-1}\, k! \, (s+\el-1)!} \Del_{\bf w}^{s+\el-1} \left[{(\langle{\bf x},{\bf w}\rangle - x_0 {\bf w})^{2s+\el-1}}\, {\bf w}^{\el-1}\right].
 \end{multline*}}
 Similarly, if $m=0$,
  \begin{multline*}
 \left( \frac{1}{\sigma_{-2n}}\int_{\Sa^{-1,2n}} \big[ \cos\left(\langle \m{w}\p, \m{x}\p \rangle \, \pa_{x_0} \right) + \m{w}\p \sin\left(\langle \m{w}\p, \m{x}\p \rangle \, \pa_{x_0} \right) \big] \, dS_{\m{w}\p} \right) x_0^{2s+\el-1} \\= (-1)^{\el-1} \frac{(n-s-\el+1)!}{4^{s+\el-1}\, n! \, (s+\el-1)!} \Del_{\m{w}\p}^{s+\el-1} \left[{(\langle{\m{x}\p},{\m{w}\p}\rangle - x_0 {\m{w}\p})^{2s+\el-1}}\, {\m{w}\p}^{\,\el-1}\right].
 \end{multline*}
\end{remark}
 
 \section{Plane wave decomposition of the Cauchy kernel}\label{PWDCKSect}
As an application of the generalized CK-extension Theorems \ref{Gen_CK_extension} and \ref{PWDecompCK}, in this section we obtain a plane wave decomposition for the fundamental solution of the Cauchy-Riemann operator $\pa_{\bf x}-\pa_{x_0}$.  First we provide a brief overview of some important properties of {this fundamental solution} and afterwards, we proceed to compute its decomposition into plane waves.


\subsection{{Fundamental solution of $\pa_{\bf x}-\pa_{x_0}$}}
In \cite{MR2386499}, fundamental solutions for the super Dirac operator $\pa_{\bf x}$ and the Laplace operator $\Del_{m|2n}$ were obtained. In particular, the fundamental solution of $\Del_{m|2n}$ was calculated to be
\begin{equation}\label{FundSol1}
\nu_2^{m|2n}=\pi^n \sum_{j=0}^{n} \frac{(-4)^j j!}{(n-j)!}\, \nu_{2j+2}^{m|0} \, \underline{x}\p^{\, 2n-2j}, 
\end{equation}
where $ \nu_{2j+2}^{m|0}$ is the fundamental solution of $\Del_{m|0}^{j+1}$ {and} $\Del_{m|0}=\sum_{j=0}^m \pa_{x_j}^2$ is the classical Laplacian with respect to the purely bosonic vector variable $\m{x}$. The superdistribution $\nu_2^{m|2n}$ satisfies 
\[\Del_{m|2n} \, \nu_2^{m|2n}({\bf x})=\del(\underline{x}) \frac{\pi^n}{n!} \underline{x}\p^{\, 2n} = \del({\bf x}).
\]
Here $\del({\bf x})=\del(\underline{x}) \frac{\pi^n}{n!} \underline{x}\p^{\, 2n} = \pi^n \del(\m{x}) x\p_1 \cdots x\p_{2n}$ is the Dirac distribution on the supervector variable ${\bf x}$ and $\del(\underline{x})=\del(x_1)\cdots \del(x_m)$ is the $m$-dimensional real Dirac distribution. Indeed, it can be verified that
\[\langle \del({\bf x}), G({\bf x})\rangle= \int_{\R^{m}} \int_B \del({\bf x})G({\bf x}) \, dV_{\m{x}}=G(0),
\]
where $G\in C^\infty(U)\otimes \mathfrak{G}_{2n}$ with $U\inc \R^{m}$ being a neighborhood of the origin.

We are now in {a position to compute} a fundamental solution for  $\pa_{\bf x}-\pa_{x_0}$. We recall that (see (\ref{LapM+1}))
\[
(\pa_{\bf x}-\pa_{x_0})(-\pa_{\bf x}-\pa_{x_0})=\Del_{m+1|2n}, \;\;\;\mbox{ and } \;\;\;(\pa_{\m{x}}-\pa_{x_0})(-\pa_{\m{x}}-\pa_{x_0})=\Del_{m+1|0}
\]
where $\Del_{m+1|0}:= \pa_{x_0}^2 + \Del_{m|0}$ is the bosonic part of $\Del_{m+1|2n}$. 
 Then a fundamental solution of $\pa_{\bf x}-\pa_{x_0}$ can be obtained by computing $(-\pa_{\bf x}-\pa_{x_0})\nu_2^{m+1|2n}$, where $\nu_2^{m+1|2n}$ is the fundamental solution of $\Del_{m+1|2n}$  given in (\ref{FundSol1}). This leads to the following result.
%
\begin{lem}
A fundamental solution of $\pa_{\bf x}-\pa_{x_0}$ is given by 
\begin{equation}\label{FundSolSupDirOp}
\fhi_1^{m+1|2n} = \pi^n \sum_{j=0}^n \frac{(-1)^j 2^{2j} j!}{(n-j)!}\, \fhi_{2j+1}^{m+1|0} \underline{x}\p^{\, 2n-2j} - \pi^n \sum_{j=0}^{n-1} \frac{(-1)^j 2^{2j+1} j!}{(n-j-1)!}\, \nu_{2j+2}^{m+1|0} \underline{x}\p^{\, 2n-2j-1},
\end{equation}
where $\fhi_{2j+1}^{m+1|0}:= (\pa_{\m{x}}-\pa_{x_0}) \nu_{2j+2}^{m+1|0}$ is a fundamental solution of the operator $(-\pa_{\m{x}}-\pa_{x_0})\Del_{m+1|0}^{j}$.
\end{lem}
%
%

In the purely bosonic case, $-\fhi_1^{m+1|0}$ is known as the Clifford Cauchy kernel since it is the fundamental solution of the generalized Cauchy-Riemann operator  $\pa_{x_0}+\pa_{\m{x}}$. This is a vector-valued kernel that can be written as the quotient $\frac{1}{\sigma_{m+1}}  \frac{x_0-\m{x}}{|x_0-\m{x}|^{m+1}}$, where $|x_0-\m{x}|= \left(x_0^2 + \sum_{j=1}^m x_j^2\right)^{1/2}$ is the Euclidean norm in $\R^{m+1}$.   For certain values of the superdimension $M$, we can still find such a form for the Cauchy kernel in the superspace setting.
 
 \begin{lem}\label{Fund_Sol_Ser}
If $M+1\notin -2\N_0$, the fundamental solution $\fhi_1^{m+1|2n}$ has the form
\[\fhi_1^{m+1|2n} = \frac{-1}{\sigma_{M+1}}  \frac{x_0-{\bf x}}{|x_0-{\bf x}|^{M+1}}.\]
{Moreover, if $|\m{x}|< |x_0|$, we obtain
\begin{equation}\label{TayCauKer}
\fhi_1^{m+1|2n}= \frac{-\sgn(x_0)}{\sigma_{M+1}} \left( \sum_{j=0}^\infty \frac{{\bf x}^{2j}}{j!} \frac{\Gam\left(\frac{M+1}{2}+j\right)}{\Gam\left(\frac{M+1}{2}\right)} |x_0|^{-M-2j} -\sgn(x_0) \sum_{j=0}^\infty \frac{{\bf x}^{2j+1}}{j!} \frac{\Gam\left(\frac{M+1}{2}+j\right)}{\Gam\left(\frac{M+1}{2}\right)} |x_0|^{-M-2j-1}\right),
\end{equation}
where $\sgn(x_0)$ {denotes} the sign of $x_0$ and $|x_0-{\bf x}|=|x_0+{\bf x}|= \left(|x_0+\m{x}|^2-\underline{x}\p^{\, 2}\right)^{\frac{1}{2}}$.}
\end{lem}
\pf 
We first recall that the fundamental solution $\nu_{2j}^{m+1|0}$ of $\Del_{m+1|0}^{j}$ is given by (see \cite{MR745128})
\begin{equation}\label{PolyMonFundSolEven}
\nu_{2j}^{m+1|0} = \frac{(-1)^j \Gam\left(\frac{m+1}{2}-j\right)}{2^{2j} \pi^{\frac{m+1}{2}} \Gam(j)} \frac{|x_0+\m{x}|^{2j}}{|x_0+\m{x}|^{m+1}}, \;\;\;\;\; \mbox{ if } \;\; m+1-2j\notin -2\N_0.
\end{equation}
This formula applies for every $j\leq n$, since the condition $M+1\notin -2\N_0$ directly implies $m+1-2j\notin -2\N_0$. 

\noindent We also recall that $(\pa_{\m{x}}-\pa_{x_0})|x_0+\m{x}|^\al= \al |x_0+\m{x}|^{\al-2} (\m{x}-x_0)$ for all $\al\in \R$. 
Thus we obtain 
\begin{align}\label{PolyMonFundSolOdd}
\fhi_{2j+1}^{m+1|0}&= (\pa_{\m{x}}-\pa_{x_0}) \nu_{2j+2}^{m+1|0} = \frac{(-1)^{j+1} \Gam\left(\frac{m+1}{2}-j\right)}{2^{2j+1} \pi^{\frac{m+1}{2}} \Gam(j+1)} \frac{(x_0-\m{x}) |x_0+\m{x}|^{2j}}{|x_0+\m{x}|^{m+1}}, & j&=0,\ldots, n-1.
\end{align}
It is easily seen that (\ref{PolyMonFundSolOdd}) also holds for $j=n$. Indeed, if we write
\[\fhi_{2n+1}^{m|0} =  \frac{(-1)^{n+1} \Gam\left(\frac{m+1}{2}-n\right)}{2^{2n+1} \pi^{\frac{m+1}{2}} \Gam(n+1)} \frac{(x_0-\m{x}) |x_0+\m{x}|^{2n}}{|x_0+\m{x}|^{m+1}},\]
we immediately obtain $(-\pa_{\underline{x}}-\pa_{x_0})\fhi_{2n+1}^{m|0}=\frac{(-1)^n \Gam\left(\frac{m+1}{2}-n\right)}{2^{2n} \pi^{\frac{m+1}{2}} \Gam(n)} \frac{|x_0+\m{x}|^{2n}}{|x_0+\m{x}|^{m+1}}=\nu_{2n}^{m+1|0}$. This means that the above expression for $\fhi_{2n+1}^{m|0}$ constitutes a fundamental solution for $(-\pa_{\m{x}}-\pa_{x_0})\Del_{m+1|0}^{n}$.

\noindent Substituting (\ref{PolyMonFundSolEven})-(\ref{PolyMonFundSolOdd}) into (\ref{FundSolSupDirOp})  we get
\begin{align}\label{IntFundSolFracForm}
\fhi_1^{m+1|2n} &= \frac{-1}{2\pi^{\frac{m+1}{2}-n}} \sum_{j=0}^n \frac{ \Gam\left(\frac{m+1}{2}-j\right)}{(n-j)!}\, \frac{(x_0-\m{x}) |x_0+\m{x}|^{2j}}{|x_0+\m{x}|^{m+1}} \underline{x}\p^{\, 2n-2j} \nonumber \\
&\phantom{=}+ \frac{1}{2\pi^{\frac{m+1}{2}-n}} \sum_{j=0}^{n-1} \frac{ \Gam\left(\frac{m+1}{2}-j-1\right)}{(n-j-1)!}\, \frac{ |x_0+\m{x}|^{2j+2}}{|x_0+\m{x}|^{m+1}} \underline{x}\p^{\, 2n-2j-1} ,
 \nonumber \\
&= \frac{-(x_0-\m{x}-\m{x}\p)}{2\pi^{\frac{M+1}{2}}}  \sum_{j=0}^{n}  \frac{ \Gam\left(\frac{m+1}{2}-j\right)}{(n-j)!}\, |x_0+\m{x}|^{2j-m-1}\underline{x}\p^{\, 2n-2j}.  
\end{align}
We now recall that $\frac{\Gam(p+1)}{\Gam(p-j+1)}=(-1)^j \frac{\Gam(-p+j)}{\Gam(-p)}$. From (\ref{GenPow}) we thus obtain
\begin{align}\label{TayExpPow}
\frac{1}{|x_0+{\bf x}|^{M+1}} &= \left(|x_0+\m{x}|^2-\underline{x}\p^{\, 2}\right)^{-\frac{M+1}{2}} 
= \frac{1}{\Gam\left(\frac{m+1}{2}-n \right)}  \sum_{j=0}^n  \frac{\Gam\left(\frac{m+1}{2}-j\right)}{(n-j)!} |x_0+\m{x}|^{2j-m-1} \, \underline{x}\p^{\, 2n-2j}.
\end{align}
Substituting {the latter} into (\ref{IntFundSolFracForm}) we obtain 
\[\fhi_1^{m+1|2n} =  \frac{-(x_0-{\bf x})}{2\pi^{\frac{M+1}{2}}} \frac{\Gam\left(\frac{M+1}{2}\right)}{|x_0+{\bf x}|^{M+1}} = \frac{-1}{\sigma_{M+1}}  \frac{x_0-{\bf x}}{|x_0-{\bf x}|^{M+1}}, \]
which is the first equality of the Lemma. 

\noindent The second equality of the Lemma follows from rewriting the Taylor expansion (\ref{TayExpPow}) of ${|x_0+{\bf x}|^{-(M+1)}}$ as a power series in the variable ${\bf x}=\m{x}+\m{x}\p$. In general, the Taylor series of the analytic function $(x_0^2+z^2)^{-\frac{M+1}{2}}$ centered at the origin converges in the circle $|z|<|x_0|$. Hence {by Lemma \ref{Lem1} we obtain} 
\begin{align*}
\frac{1}{|x_0+{\bf x}|^{M+1}} &= \left(x_0^2-{\bf x}^2\right)^{-\frac{M+1}{2}}  
= \sum_{j=0}^\infty  \frac{{\bf x}^{2j}}{j!} \frac{\Gam\left(\frac{M+1}{2}+j\right)}{\Gam\left(\frac{M+1}{2}\right)} |x_0|^{-M-2j-1}, & \mbox{ if } \;\;& |\m{x}|<|x_0|.
\end{align*}
Multiplying the above identity by $x_0-{\bf x}$ yields the result. $\hfill\square$

\subsection{{Plane wave decomposition}}
Lemma \ref{Fund_Sol_Ser} assures that, if $M+1\notin -2\N_0$, $x_0\neq 0$ {and $|\m{x}|<|x_0|$,} then $\fhi_1^{m+1|2n}(x_0,{\bf x})$ is a monogenic power series in ${\bf x}$ and thus, a generalized CK-extension of certain initial functions of the real variable $x_0$. {Now we} proceed to explicitly obtain plane wave decompositions for this Cauchy kernel with $M+1\notin -2\N_0$. We shall treat separately the cases {where} $M\notin -2\N_0$ (i.e.\ $M\geq1$), $M=-2k$ with $m\neq 0$, and the case $m=0$.

\paragraph{Case $M \geq 1$.} By Corollary \ref{PWCKx_0} $i)$ we have
\[\frac{-1}{\sigma_{M+1}}  \frac{x_0-{\bf x}}{|x_0-{\bf x}|^{M+1}} = \left(\frac{1}{\sigma_M} \int_{\Sa^{m-1,2n}} \exp\left(\langle {\bf w}, {\bf x}\rangle \, {\bf w}  \, \pa_{x_0}\right) \, dS_{\bf w} \right) F_0({x_0}), \]
where the initial function $F_0$ is given in this case  by $F_0(x_0)= \frac{-\sgn(x_0)}{\sigma_{M+1}} |x_0|^{-M}$, see (\ref{TayCauKer}). Consider now the complex function $g(z)=z^{-M}$, which is holomorphic in $\C\setminus\{0\}$. The real and imaginary parts of $g(ix_0)$ are respectively given by 
\begin{align*}
g_1(0,x_0)&= \begin{cases} (-1)^{\frac{M}{2}} x_0^{-M}, & M \mbox{ even}, \\ 0, & M \mbox{ odd}, \end{cases} & 
g_2(0,x_0)&= \begin{cases} 0, & M \mbox{ even}, \\ (-1)^{\frac{M+1}{2}} x_0^{-M}, & M \mbox{ odd}. \end{cases}
\end{align*}
Then, for $M$ even we can write $F_0(x_0)= \frac{-(-1)^{\frac{M}{2}} \sgn(x_0)}{\sigma_{M+1}} g_1(0,x_0)$, and from Lemma \ref{Int_PW} we obtain
\begin{align}
\frac{-1}{\sigma_{M+1}}  \frac{x_0-{\bf x}}{|x_0-{\bf x}|^{M+1}} &= \frac{-(-1)^{\frac{M}{2}} \sgn(x_0)}{\sigma_{M+1} \sigma_{M}} \left( \int_{\Sa^{m-1,2n}} \exp\left(\langle {\bf w}, {\bf x}\rangle \, {\bf w}  \, \pa_{x_0}\right) \, dS_{\bf w} \right)g_1(0,x_0) \nonumber \\
&= -\sgn(x_0) \frac{(-1)^{\frac{M}{2}}(M-1)!}{2 (2\pi)^M} \int_{\Sa^{m-1,2n}} (\langle{\bf x},{\bf w}\rangle - x_0 {\bf w})^{-M}\, dS_{\bf w}, \label{PWDM1E}
\end{align}
where {the identity $\sigma_{M+1} \sigma_{M} = \frac{2 (2\pi)^M}{(M-1)!}$ has been used.}

\noindent For $M$ odd we have $F_0(x_0)= \frac{-(-1)^{\frac{M+1}{2}} }{\sigma_{M+1}} g_2(0,x_0)$ and therefore
\begin{align}
\frac{-1}{\sigma_{M+1}}  \frac{x_0-{\bf x}}{|x_0-{\bf x}|^{M+1}} &= \frac{-(-1)^{\frac{M+1}{2}}}{\sigma_{M+1} \sigma_{M}} \left( \int_{\Sa^{m-1,2n}} \exp\left(\langle {\bf w}, {\bf x}\rangle \, {\bf w}  \, \pa_{x_0}\right) \, dS_{\bf w} \right)g_2(0,x_0) \nonumber\\
&= - \frac{(-1)^{\frac{M+1}{2}}(M-1)!}{2 (2\pi)^M} \int_{\Sa^{m-1,2n}} (\langle{\bf x},{\bf w}\rangle - x_0 {\bf w})^{-M} {\bf w}\, dS_{\bf w}. \label{PWDM1O}
\end{align}

\noindent {The proofs of (\ref{PWDM1E})  and (\ref{PWDM1O}) make use of the power series (\ref{TayCauKer}), and therefore, they depend on the assumption that $|\m{x}|<|x_0|$. However, this restriction may be easily removed due to the uniqueness of analytic continuation. Indeed, since $x_0\neq 0$, both sides in (\ref{PWDM1E})  and (\ref{PWDM1O}) are given by real analytic (super) functions on $\m{x}\in\R^m$. Since the functions in both sides of these formulas respectively coincide in the neighborhood $|\m{x}|<|x_0|$ of the origin, they must be identical everywhere in $\R^m$.
}

\noindent In this way, we have obtained plane wave decompositions for $\fhi_1^{m+1|2n}$ if $M\geq 1$. {These formulas} resemble the same structure {of their analogues (\ref{PWCKTrad})} in the purely bosonic case, see also \cite{MR1169463, MR985370}.
 As we shall see next, the cases of negative superdimension will bring new structures into {these plane wave decompositions.} 

\paragraph{Case $M =-2k$ and $m\neq 0$.} From Corollary \ref{PWCKx_0} $i)$ we obtain
\begin{multline}\label{PWCauchy2}
\frac{-1}{\sigma_{M+1}}  \frac{x_0-{\bf x}}{|x_0-{\bf x}|^{M+1}} = \left(\frac{1}{\sigma_{-2k}} \int_{\Sa^{m-1,2n}} \exp\left(\langle {\bf w}, {\bf x}\rangle \, {\bf w}  \, \pa_{x_0}\right) \, dS_{\bf w} \right) F_0(x_0)\\
 + k! (4\pi)^k \left( \int_{\Sa^{m-1,2n}} \exp\left(\langle {\bf w}, {\bf x}\rangle \, {\bf w}  \, \pa_{x_0}\right) \, dS_{\bf w} \right) B({x_0}),
\end{multline}
where $F_0(x_0)= \frac{-\sgn(x_0)}{\sigma_{-2k+1}} x_0^{2k}$, $F_{2k+1}(x_0)= \frac{\pi^k}{2 \, k!} |x_0|^{-1}$ and $\pa_{x_0}^{2k+1} B = F_{2k+1}$. {Application of Lemma 7 $i)$ for $s = k$ and $\el= 1$ enables us to compute the first normalized integral of (\ref{PWCauchy2}), denoted by $I_1$, as follows (see also Remark \ref{RemPW})}
 \begin{align*}
 I_1 &=  \frac{-  \sgn(x_0)}{\sigma_{-2k+1}}  \left(\frac{1}{\sigma_{-2k}} \int_{\Sa^{m-1,2n}} \exp\left(\langle {\bf w}, {\bf x}\rangle \, {\bf w}  \, \pa_{x_0}\right) \, dS_{\bf w} \right){x_0^{2k}} \\
 &= \frac{- (-1)^k \sgn(x_0)}{\sigma_{-2k+1}} \frac{1}{\sigma_{-2k}} \int_{\Sa^{m-1,2n}} (\langle{\bf x},{\bf w}\rangle - x_0 {\bf w})^{2k} \, dS_{\bf w} \\
  &= \frac{- \sgn(x_0)}{ 2^{2k} (k!)^2 \;\sigma_{-2k+1}}\; \Del_{\bf w}^k (\langle{\bf x},{\bf w}\rangle - x_0 {\bf w})^{2k}.
 \end{align*}

\noindent In order to compute the second integral of (\ref{PWCauchy2}), {denoted by  $I_2$}, we must find first a good candidate for the function $B(x_0)$. To that end, we consider the sequence of functions
\begin{equation}\label{RecFunc}
G_\el(z) = \frac{z^\el}{\el!} \ln(z) - a_\el z^\el, \;\;\;\; \mbox{ with } \;\;\;  a_{\el+1}=\frac{1}{\el+1} \left(a_\el + \frac{1}{(\el+1)!}\right), \;\; a_0=0,
\end{equation}
defined in the principle branch $-\pi < Arg(z) \leq \pi$ of the logarithm function in the complex plane. 
The sequence $\{a_\el\}$ can be explicitly redefined as $a_\el = \frac{\Psi(\el+1) - \Psi(1)}{\el!}$ where $\Psi(z) = \frac{\Gam'(z)}{\Gam(z)}$ is the digamma function.

\noindent It is easily seen that $G'_{\el+1}(z)=G_\el(z)$ with $G_0(z)=\ln(z)$, and therefore, $G_{2k}^{(2k+1)}(z)=z^{-1}$. Then, if we choose \[B(x_0)=  \frac{\pi^k}{2\, k!} \sgn(x_0) G_{2k}(|x_0|),\] we obtain $\pa_{x_0}^{2k+1} B = F_{2k+1}$. Moreover, the real part of $G_{2k}(ix_0)$ can be computed to be 
\[G_{2k,1}(0,x_0) = (-1)^k \left(\frac{x_0^{2k}}{(2k)!} \ln(|x_0|) - a_{2k} x_0^{2k}\right) =(-1)^k G_{2k}(|x_0|).\]
Hence, {$I_2$ can be rewritten as (see Lemma \ref{Int_PW})}
\begin{align*}
I_2 &= \left( \int_{\Sa^{m-1,2n}} \exp\left(\langle {\bf w}, {\bf x}\rangle \, {\bf w}  \, \pa_{x_0}\right) \, dS_{\bf w} \right) B({x_0})\\
&=  \frac{(-1)^k  \,\pi^k}{2\, k!} \sgn(x_0) \left( \int_{\Sa^{m-1,2n}} \exp\left(\langle {\bf w}, {\bf x}\rangle \, {\bf w}  \, \pa_{x_0}\right) \, dS_{\bf w} \right)G_{2k,1}(0,x_0) \\
&= \frac{(-1)^k  \,\pi^k}{2\, k!} \sgn(x_0)  \int_{\Sa^{m-1,2n}} G_{2k}\left(\langle{\bf x},{\bf w}\rangle - x_0 {\bf w}\right) \, dS_{\bf w}.
\end{align*}
Finally, {the following decomposition holds}
\begin{multline}\label{PWDCKM2k}
\frac{-1}{\sigma_{M+1}}  \frac{x_0-{\bf x}}{|x_0-{\bf x}|^{M+1}} = \frac{- \sgn(x_0)}{ 2^{2k} (k!)^2 \;\sigma_{-2k+1}}\; \Del_{\bf w}^k (\langle{\bf x},{\bf w}\rangle - x_0 {\bf w})^{2k}\\
 + \frac{(-1)^k (4\pi^2)^k}{2} \sgn(x_0) \int_{\Sa^{m-1,2n}} G_{2k}\left(\langle{\bf x},{\bf w}\rangle - x_0 {\bf w}\right) \, dS_{\bf w}.
\end{multline}
Similarly to the previous case, the proof of (\ref{PWDCKM2k}) uses the convergence of the power series (\ref{TayCauKer}) and of the series (\ref{Taylor_PW}) for the function $G_{2k}\left(\langle{\bf x},{\bf w}\rangle - x_0 {\bf w}\right)$. Both series are convergent if $|\m{x}|<|x_0|$, but the identification (\ref{PWDCKM2k}) holds everywhere due to the uniqueness of analytic continuation.

\paragraph{Case $M =-2n$ ($m= 0$).} In this case, the Cauchy kernel is represented by a finite power series in the fermionic vector variable $\m{x}\p$, see Lemma \ref{Fund_Sol_Ser}. From Corollary \ref{PWCKx_0} $iii)$ we thus obtain 
\[
\frac{-1}{\sigma_{-2n+1}}  \frac{x_0-{\m{x}\p}}{|x_0-\m{x}\p|^{-2n+1}} =  \left(\frac{1}{\sigma_{-2n}} \int_{\Sa^{-1,2n}} \big[ \cos\left(\langle \m{w}\p, \m{x}\p \rangle \, \pa_{x_0} \right) + \m{w}\p \, \sin\left(\langle \m{w}\p, \m{x}\p \rangle \, \pa_{x_0} \right) \big] \, dS_{\m{w}\p} \right) F_0({x_0}),\]
with $F_0({x_0}) = \frac{-\sgn(x_0)}{\sigma_{-2n+1}} x_0^{2n}$. Then, using Lemma \ref{NorInt_PW} $ii)$ we obtain (see also Remark \ref{RemPW})
\begin{align*}
\frac{-1}{\sigma_{-2n+1}}  \frac{x_0-{\m{x}\p}}{|x_0-\m{x}\p|^{-2n+1}}  &=  \frac{-\sgn(x_0)}{\sigma_{-2n+1}}  \left(\frac{1}{\sigma_{-2n}} \int_{\Sa^{-1,2n}} \big[ \cos\left(\langle \m{w}\p, \m{x}\p \rangle \, \pa_{x_0} \right) + \m{w}\p \, \sin\left(\langle \m{w}\p, \m{x}\p \rangle \, \pa_{x_0} \right) \big] \, dS_{\m{w}\p} \right) x_0^{2n} \\
&=  \frac{-(-1)^n \, \sgn(x_0)}{\sigma_{-2n+1}} \;  \frac{1}{\sigma_{-2n}} \int_{\Sa^{-1,2n}} (\langle\m{x}\p,\m{w}\p\rangle - x_0 \m{w}\p)^{2n} dS_{\m{w}\p} \\
&= \frac{-\sgn(x_0)}{4^n \, (n!)^2 \, \sigma_{-2n+1}}\, \Del_{\m{w}\p}^n \left[(\langle\m{x}\p,\m{w}\p\rangle - x_0 \m{w}\p)^{2s}\right].
\end{align*}

Summarizing, we have obtained the following plane wave decompositions for the Cauchy kernel in superspace when $M+1\notin -2\N_0$.
\begin{teo}\label{PWDCK}
Let $x_0\neq 0$ and $M+1\notin -2\N_0$. Then
\begin{itemize}
\item[$i)$] If $M>1$,
\begin{align*}
\frac{-1}{\sigma_{M+1}}  \frac{x_0-{\bf x}}{|x_0-{\bf x}|^{M+1}} &= -\sgn(x_0) \frac{(-1)^{\frac{M}{2}}(M-1)!}{2 (2\pi)^M} \int_{\Sa^{m-1,2n}} (\langle{\bf x},{\bf w}\rangle - x_0 {\bf w})^{-M}\, dS_{\bf w}, & \mbox{ for }& \;\;M \mbox{ even},\\[+.2cm]
\frac{-1}{\sigma_{M+1}}  \frac{x_0-{\bf x}}{|x_0-{\bf x}|^{M+1}}&= - \frac{(-1)^{\frac{M+1}{2}}(M-1)!}{2 (2\pi)^M} \int_{\Sa^{m-1,2n}} (\langle{\bf x},{\bf w}\rangle - x_0 {\bf w})^{-M} {\bf w}\, dS_{\bf w}, & \mbox{ for }& \;\;M \mbox{ odd}.
\end{align*}

\item[$ii)$] If $M =-2k$ and $m\neq 0$,
\begin{multline*}
\frac{-1}{\sigma_{M+1}}  \frac{x_0-{\bf x}}{|x_0-{\bf x}|^{M+1}} = \frac{- \sgn(x_0)}{ 2^{2k} (k!)^2 \;\sigma_{-2k+1}}\; \Del_{\bf w}^k (\langle{\bf x},{\bf w}\rangle - x_0 {\bf w})^{2k}\\
 + \frac{(-1)^k (4\pi^2)^k}{2} \sgn(x_0) \int_{\Sa^{m-1,2n}} G_{2k}\left(\langle{\bf x},{\bf w}\rangle - x_0 {\bf w}\right) \, dS_{\bf w},
\end{multline*}
with $G_{2k}$ defined as in (\ref{RecFunc}).

\item[$iii)$] If $M =-2n$ ($m= 0$),
\begin{align*}
\frac{-1}{\sigma_{-2n+1}}  \frac{x_0-{\m{x}\p}}{|x_0-\m{x}\p|^{-2n+1}} &= \frac{-\sgn(x_0)}{4^n \, (n!)^2 \, \sigma_{-2n+1}}\, \Del_{\m{w}\p}^n \left[(\langle\m{x}\p,\m{w}\p\rangle - x_0 \m{w}\p)^{2n}\right].
\end{align*}
\end{itemize}
\end{teo}

\section*{Acknowledgements}
The author gratefully acknowledges the many helpful suggestions of Frank Sommen, Irene Sabadini, Juan Bory-Reyes, Michael Wutzig and Hendrik De Bie during the preparation of the paper.  The author is supported by a B.O.F. postdoctoral grant from Ghent University with grant number BOF18/PDO/073.

\bibliographystyle{abbrv}

\end{document}